\documentclass{emulateapj}

\newcommand{\kms}{km~s$^{-1}$}
\newcommand{\msunyr}{$M_\sun$~yr$^{-1}$}
\newcommand{\lsun}{$L_\sun$}
\newcommand{\bgamma}{Br$\gamma$}
\newcommand{\bten}{Br${10}$}
\newcommand{\pgamma}{Pa$\gamma$}
\newcommand{\pbeta}{Pa$\beta$}
\newcommand{\pdelta}{Pa$\delta$}

\slugcomment{Accepted by ApJ}

\shorttitle{NIR H Lines in T Tauri Stars}
\shortauthors{Edwards et al.}

\begin{document}

\title{Interpreting Near Infrared Hydrogen Line Ratios in T Tauri Stars}

\author{Suzan Edwards\altaffilmark{1,5}, John Kwan\altaffilmark{2}, William Fischer\altaffilmark{3,5}, Lynne Hillenbrand\altaffilmark{4}, Kimberly Finn\altaffilmark{6}, Kristina Fedorenko\altaffilmark{1}, and Wanda Feng\altaffilmark{1}}
\altaffiltext{1}{Five College Astronomy Department, Smith College, Northampton, MA 01063, sedwards@smith.edu}
\altaffiltext{2}{Five College Astronomy Department, University of Massachusetts, Amherst, MA 01003, kwan@astro.umass.edu}
\altaffiltext{3}{Department of Physics and Astronomy, University of Toledo, Toledo, OH 43606, wfische@utnet.utoledo.edu}
\altaffiltext{4}{Department of Astronomy, California Institute of Technology, Pasadena, CA 91125, lah@astro.caltech.edu}
\altaffiltext{5}{Visiting Astronomer, NASA Infrared Telescope Facility}
\altaffiltext{6}{Five College Astronomy Department, Mount Holyoke College, South Hadley, MA 01063}

\begin{abstract}
In accreting young stars one of the prominent spectral features in the near infrared is the Paschen and Brackett series in emission. We examine hydrogen line ratios for 16 classical T Tauri stars from SpeX spectra and and assess the trends with veiling and accretion. The observed line ratios are compared to two theoretical models for line formation:\ (1) Baker and Menzel's (1938) Case B for radiative ionization and recombination and (2) a set of local line excitation calculations designed to replicate the conditions in T Tauri winds and magnetic accretion columns \citep{kwan11}. While the comparison between Case B and observed line ratios implies a wide range in electron density and temperature among the hydrogen line formation regions in T Tauri stars,  the predictions of the local line excitation models give consistent results across multiple diagnostics.  Under the assumptions of the local line excitation calculations, we find that $n_H$ in the hydrogen line formation region is constrained to $2 \times 10^{10} - 2 \times 10^{11}$ cm$^{-3}$, where stars with higher accretion rates have densities at the higher end of this range. Because of uncertainties in extinction, temperature is not well delineated but falls within the range expected for collisional excitation to produce the line photons. We introduce new diagnostics for assessing extinction based on near infrared hydrogen line ratios from the local line excitation calculations.
\end{abstract}


\section{Introduction}

Some of the most prominent emission lines in T Tauri stars are from hydrogen. In the first few decades of T Tauri research the Balmer emission lines were attributed to formation in energetic winds \citep{kuhi64,har82}, but in the mid-nineties a reassessment of the hydrogen emission line profile morphology in the context of magnetospheric infall \citep{cal92, har94} became the underpinning for the current paradigm that these young stars are in the final stages of accretion from a protoplanetary disk.  Early successes of the model included generating Balmer emission line profiles that are centrally peaked and have a small blue asymmetry, with inverse P Cygni profiles at favorable inclinations \citep{muz98a,edw94}. A series of papers culminating in \citet{muz01} carried out a grid of radiative transfer models for line formation under the assumption of magnetospheric accretion in a dipole field geometry, finding good agreement in general profile morphology and line luminosity for hydrogen lines in many stars, including the infrared lines {\pbeta} and {\bgamma}. 

These models have more recently been incorporated into composite accretion and disk wind scenarios \citep{lima} and into time dependent 3-D numerical simulations of magnetospheric accretion, including both dipole and multipole configurations, which may be aligned or misaligned with the stellar rotation axis \citep{kur08}, and accompanied by MHD winds from the inner disk \citep{kur12,kur11}. In general the magnetospheric accretion models do a reasonable job of describing the morphology of the hydrogen line profiles and their luminosities, and are widely accepted as the origin of hydrogen emission in classical T Tauri (CTTS) stars. However, shortcomings of these models have been identified by \citet{ale00}, \citet{bek} and \citet{kur11}, based on comparing predictions and observations of line profile morphology, especially in the larger extent of the blue wing emission in observed compared to model profiles. 

Regardless of their origin, several groups have found well defined relations between the hydrogen line luminosity and the disk accretion rate, where the latter is assessed from accretion shock models of the excess optical and ultraviolet emission attributed to the post-shock heated photosphere at the magnetospheric footprints, that also includes contributions from the pre-shock gas \citep{cal98}. This empirical correlation between hydrogen  line luminosity and disk accretion rate, first established for {\pbeta} and {\bgamma} by \citet{muz98c}, has become a powerful means of determining disk accretion rates for YSOs over a wide range of masses included embedded objects, where high extinction prevents optical and ultraviolet emission excess above the photosphere from being observed and modeled \citep{cal04, nat06, gatti06, her08,rig12}. 

Improved understanding of hydrogen emission in accreting stars will require a confrontation between the physical conditions required to produce the line luminosities and line ratios in combination with high resolution profile studies that define the kinematics of the line formation region. To date, the hydrogen line luminosities in the magnetospheric accretion models are treated in an ad-hoc way, where accretion rates through the funnel flow set the density of infalling gas, and line luminosities require fine-tuning the temperature and size of the flow, with larger sizes and cooler temperatures in the higher accretion rate, higher line luminosity sources. Required temperatures range from lows of 6000 -- 7000~K for high accretion rate sources to 10,000 -- 20,000 K for low accretion rate sources \citep{muz98a}.   Attempts to self consistently assess heating and cooling in funnel flows are more restrictive in allowed temperatures \citep{martin96} and the inferred low temperatures cannot account for the observed hydrogen line luminosities \citep{muz98a}.  However, this ad-hoc approach to determining the temperature in a funnel flow does result in a relation between line luminosity and accretion rate similar to what is observed by assessing accretion rates from excess continuum emission, providing an overall consistency with this interpretation of the hydrogen emission. 

A more direct approach for assessing the physical conditions in the hydrogen line formation region is to use observed line luminosities and line ratios as direct diagnostics of the density and temperature based on calculations of atomic level populations. Natta, Giovanardi, and Palla took this approach in their 1988 paper, exploring the ionization and excitation structure of hydrogen lines formed in dense, cool winds from low luminosity pre-main sequence stars, with mass loss rates between $10^{-8}$ and $10^{-6}$ \msunyr. They found that line luminosities increase with mass loss rate, although different assumptions for the gas temperature and stellar radiation field produced a wide range of fluxes at a given mass loss rate. They also pointed out that line ratios in the infrared may be able to discriminate among different models.

The advantage of using near infrared hydrogen lines as diagnostics of physical conditions is significant. The line opacities are much smaller than the Balmer lines, so they rarely show blueshifted absorption from a wind and have a lower frequency of redshifted absorption from infalling gas \citep{folha, edw06, fis08} and extinction corrections are also smaller.  Historically, the most widely examined ratio in near infrared lines is {\pbeta}$/${\bgamma}. \citet{muz98a,muz98b} found this ratio, coupled with the line luminosities, to be roughly consistent with the magnetospheric models described above for 19 CTTS in Tau-Aur.  In the magnetospheric accretion scenario both lines are optically thick and {\pbeta}$/${\bgamma} ratios are between 3 and 6. A study of CTTS and brown dwarfs in $\rho$ Oph found most of the CTTS to have similar ratios to Tau-Aur but the accreting brown dwarfs and a few CTTS showed {\pbeta}$/${\bgamma} ratios $\sim$ 2, which were posited to result from low temperature, high optical depths spots in the shock heated photosphere \citep{gatti06}.  A recent investigation of this ratio in 47 sources in the Chameleon I and II star formation region \citep{ant} found that while many have ratios similar to those in Tau-Aur, almost half have ratios $\le$ 2, again posited to form in very optically thick regions with $T < 4000$ K. 

A growing number of near infrared spectrographs permit a broad range of ratios from Paschen and Brackett series decrements to be determined simultaneously, which is crucial since the lines are quite variable. A number of recent studies have compared these decrements to predictions of Baker and Menzel's Case B for radiative ionization and recombination, which are available in an online database \citep{B}. In this recombination scenario the near infrared lines are optically thin, the level populations are dominated by radiative cascade from the continuum, and collisional effects are included. This approach has yielded some surprising results. In a study of 15 stars from Tau-Aur,  \citet{bary08} found Paschen and Brackett decrements, taken as an average over all stars, to be best matched by Case B conditions with $T< 2000$ K and $n_e$ $\sim 10^{10}$~cm$^{-3}$.  Similar approaches for other individual stars indicate quite different conditions. For TW Hya, \citet{vac} find series decrements matching Case B for $T =$ 20,000 K and $n_e$ $\sim 10^{13}$ cm$^{-3}$, while \citet{podio} found reasonable agreement for Case B predictions of $T =$ 10,000~K and $n_e$ between $10^{3}$ and $10^{7}$ cm$^{-3}$ from the Brackett decrement for RU Lup. Similar approaches have been taken for other YSOs, where \citet{kos} find decrements implying $T =$ 10,000 K and $n_e$ $\sim 10^{7}$ cm$^{-3}$ for the outbursting CTTS EX Lup and \citet{kraus} find $T=$ 10,000 K and $n_e$ = $6 \times 10^{13}$ cm$^{-3}$ for the AeBe star V921 Sco. This very diverse range of physical conditions inferred from comparing Case B predictions to series decrements, with a range of over 6 orders of magnitude in density and temperatures from 1000 to 20,000~K, makes it questionable as to whether this scenario may be an appropriate choice for evaluating the physical conditions in the hydrogen line formation region in accreting young stars.  

Spectacular new instrumentation, such as the VLT's X-SHOOTER and CRIRES, ensure that abundant near infrared spectra of star forming regions will be forthcoming and it is thus of interest to identify good diagnostics for interpreting these results.  A new set of local line excitation calculations by \citet{kwan11} for physical conditions appropriate for winds and accretion flows in CTTS offers an opportunity to evaluate hydrogen line ratios in series decrements and across series without the restrictions imposed by Case B assumptions and over a wider range of physical conditions and atomic properties than in the earlier work of \citet{natta88}. Since these calculations, which evaluate line emissivities as a function of density, temperature, and ionizing flux, also include transitions of \ion{He}{1}, \ion{O}{1}, \ion{Ca}{2} and \ion{Na}{1}, they offer the opportunity to evaluate physical conditions from many emission line ratios simultaneously, presenting more rigorous tests of the line formation region than previously available.  

In this paper we look at various diagnostics from the Paschen and Brackett series for 16 T Tauri stars and compare them to predictions for both Case B and the Kwan \& Fischer local line excitation models. The sample overlaps with those of \citet{muz98a} and \citet{bary08} but the lines are resolved, allowing us to specify velocity limits in the Paschen and Brackett lines of each star that are not affected by redshifted absorption features, thus giving more precise line ratios of the emitting gas than lower resolution studies and also to make good assessments of the continuum emission excess. The presentation includes \S\ 2 describing the sample and uncertainties in extinction, \S\ 3 presenting the observed hydrogen line ratios and introducing several diagnostics to compare to model predictions, \S\ 4 comparing the assumptions for Case B and the Kwan \& Fischer local line excitation calculations, and \S\ 5 comparing model predictions to observations and examining the role of extinction uncertainties. We end with a discussion and conclusions in \S\ 6 and \S\ 7.

\section{The Sample: Line Equivalent Widths, Extinctions, Mass Accretion Rates}

Our sample is 16 classical T Tauri stars (CTTS) observed with SpeX at the Infrared Telescope Facility on 2006 November 26 and 27. It is largely drawn from the Tau-Aur star forming region, with spectral types from G8 to M2, and selected to cover a broad range of emission excess and disk accretion rates. The spectra were taken in the short-wavelength cross-dispersed (SXD) mode, with an 0.3$\arcsec$ by 15$\arcsec$ slit and a spectral extent from 0.8 to 2.4~$\mu$m at a resolving power $R=2000$. Total exposure times ranged from 16 to 48 minutes for the program stars, with  $8.2\leq J \leq10.7$, yielding a continuum $S/N \sim 250$ at $J$, increasing to longer wavelengths. The only unresolved binary in the sample is DF Tau, separation of 0.09\arcsec, of spectral types M2.0 and M2.5 with a flux ratio at $K$ of 1.62 \citep{har03, whi01}.  These spectra were included in the study of  Fischer et al.\ (2011), hereafter \citealt{fis11}, and further properties of the sample are detailed there, along with specifics of the reduction and analysis of the spectra. 

In the 2011 paper these spectra were used in conjunction with near simultaneous spectra from Keck I's HIRES and Keck II's NIRSPEC to derive continuum veiling and continuum emission excesses from 0.48 to 2.4~$\micron$. Here we focus on the near infrared hydrogen emission lines in the SpeX spectra, which were shown by \citealt{fis11} to have equivalent widths proportional to the excess continuum emission. As in previous studies, the {\it veiling} is defined as the ratio of excess to photospheric emission at a specific wavelength.  

Table~\ref{t.sample} identifies the sample, along with the literature spectral types used in FEHK, the veiling at 1 $\mu$m, $r_Y$, derived in FEHK, and the emission equivalent widths for \pbeta, \pgamma, and {\bgamma} measured from the SpeX spectra. The 1 $\mu$m veiling ranges from  just barely detectable (0.1) to quite significant (3.4) and the equivalent widths range from lows of a few tenths of an \AA\ to as high as 28 \AA\  for {\pbeta} and 10 \AA\ for {\bgamma}.  The error in equivalent width depends on the signal to noise in the continuum, the line to continuum ratio, and the width of the line (several hundred km s$^{-1}$). Most stars have equivalent width errors $\sim10-15$\% but for the three stars (AA Tau, BM And, LkCa 8) with the smallest line/continuum ratios (1.02 to 1.05 at {\bgamma}) errors can exceed 50\%. 

\begin{deluxetable}{lccrrr}
\tablecaption{SpeX CTTS Sample \label{t.sample}}
\tabletypesize{\footnotesize}
\tablewidth{\hsize}
\tablehead{\colhead{Object} & \colhead{Sp.\ Type} &\colhead{r$_{Y}$} &  \colhead{EW {\pbeta}}  &\colhead{EW \pgamma} & \colhead{EW {\bgamma}} \\ \colhead{(1)} & \colhead{(2)} & \colhead{(3)} & \colhead{(4)} & \colhead{(5)} & \colhead{(6)} }
\startdata
AA Tau	&	K7	&	0.2	&	0.3 	&	0.8 	& 1.2   \\
AS 353A	&	K5	&	2.2	&	28 	&	15   	&18  \\
BM And	&	G8	&	0.1	&	1.0  	&	0.2 	& 0.3  \\
BP Tau	&	K7	&	0.4	&	9.6  	&	6.5 	& 4.4  \\
CW Tau	&	K3	&	1.2	&	8.0  	&	4.9 	& 3.3  \\
CY Tau	&	M1	&	0.2	&	0.9 	&	1.0 	& 1.0  \\
DF Tau	&	M2	&	0.5	&	5.0  	&	3.8 	& 3.3 \\
DG Tau	&	K7	&	0.7	&	14 	&	8.9 	& 7.6  \\
DK Tau	&	K7	&	0.6	&	4.0 	&	4.2 	& 2.5  \\
DL Tau	&	K7	&	1.8	&	23 	&	16 	& 12 \\
DO Tau	&	M0	&	0.9	&	8.7 	&	6.3 	& 3.4  \\
DR Tau	&	K7	&	3.4	&	24 	&	18 	& 8.9 \\
HN Tau	&	K5	&	1.0	&	12 	&	7.3 	& 5.0 \\
LkCa 8	&	M0	&	0.2	&	0.3 	&	1.0 	& 1.0  \\
RW Aur	&	K1	&	2.1	&	26 	&	14 	 & 10 \\
UY Aur	&	M0	&	0.6	&	3.4 	&	3.9 	  & 2.0
\enddata
\tablecomments{Cols.~2, 3:\ Spectral type and 1 $\mu$m veiling from FEHK,  Cols.~4--6:\ Emission equivalent widths in \AA\ from direct SpeX spectra. Typical errors are 10--15\% except for CY Tau where the error in {\bgamma} is $\sim$30 \% and AA Tau, BM And, and LkCa 8 where errors are $\sim$30\% in  {\pbeta} and {\pgamma} and $\sim$50\% in {\bgamma}.}
\end{deluxetable}

\begin{deluxetable*}{lccccccc}
\tablecaption{Extinctions, Accretion Rates \label{t.Av}}
\tabletypesize{\footnotesize}
\tablewidth{\hsize}
\tablehead{\colhead{Object} & \colhead{A$_{V}$FEHK} &\colhead{A$_{V}$MIN} &  \colhead{A$_{V}$MAX} & \colhead{A$_{V}$MEAN}  & \colhead{A$_{V}$KF} & \colhead{log L$_{acc}$} & \colhead{log ${\dot M}_{acc} $} \\
\colhead{(1)} & \colhead{(2)} & \colhead{(3)} & \colhead{(4)} & \colhead{(5)} & \colhead{(6)} & \colhead{(7)} & \colhead{(8)} }
\startdata
AA Tau	&	1.3	&	0.5	&	2.3	&	1.2	& \nodata	&	-1.99	&	-8.87		\\
AS 353A	&	2.1	&	2.1	&	3.4	&	2.5	&	2.9	&	 0.05	&	-6.94		\\
BM And	&	1.6	&	0.7	&	1.6	&	1.3	& \nodata	&	-1.25	&	-8.61		\\
BP Tau	&	1.8	&	0.5	&	1.8	&	1.0	&	0.9	&	-0.59	&	-7.60		\\
CW Tau	&	2.1	&	2.0	&	2.4	&	2.4	&	3.0	&	-0.74	&	-8.00		\\
CY Tau	&	1.2	&	0.1	&	1.7	&	0.7	&	0.0	&	-1.97	&	-9.03		\\
DF Tau	&	1.8	&	0.2	&	2.3	&	1.1	&	1.0	&	-0.50	&	-6.88		\\
DG Tau	&	3.9	&	1.0	&	3.9	&	2.4	&	3.5	&	-0.15	&	-7.15		\\
DK Tau	&	1.8	&	0.8	&	3.1	&	1.6	&	0.3	&	-0.78	&	-7.72		\\
DL Tau	&	3.0	&	1.4	&	3.0	&	2.1	&	1.4	&	-0.34	&	-7.20		\\
DO Tau	&	3.0	&	2.0	&	3.5	&	3.0	&	1.5	&	-0.62	&	-7.37		\\
DR Tau	&	1.5	&	1.0	&	1.7	&	1.7	&	0.7	&	-0.02	&	-6.95		\\
HN Tau	&	3.1	&	0.4	&	3.1	&	1.3	&	3.0	&	-1.11	&	-8.34		\\
LkCa 8	&	0.5	&	0.2	&	2.0	&	0.7	& \nodata	&	-2.37	&	-9.33		\\
RW Aur	&	2.2	&	0.5	&	1.2	&	1.1	&	1.9	&	0.16	&	-7.05		\\
UY Aur	&	1.5	&	0.6	&	3.1	&	1.5	&	0.3	&	-0.87	&	-7.59		
\enddata
\tablecomments{Cols.~3--5:\ minimum, maximum, and mean A$_{V}$ from the literature (\citealt{fur11,bric02, gul98, gul00, ken95,val93} for Tau-Aur; \citealt{eis90}  for AS 353A; \citealt{ros99,gue93} for BM And). Col 6:\ A$_{V}$ from this paper based on KF models assuming $T=$ 10,000 K, Cols.~7, 8:\ accretion luminosity (\lsun ) and mass accretion rate (\msunyr) based on {\pbeta} equivalent width, 2MASS continuum fluxes,  FEHK A$_{V}$, and the accretion calibration of \citet{natta04}.} 
\end{deluxetable*}

\begin{figure}
\includegraphics[width=\hsize] {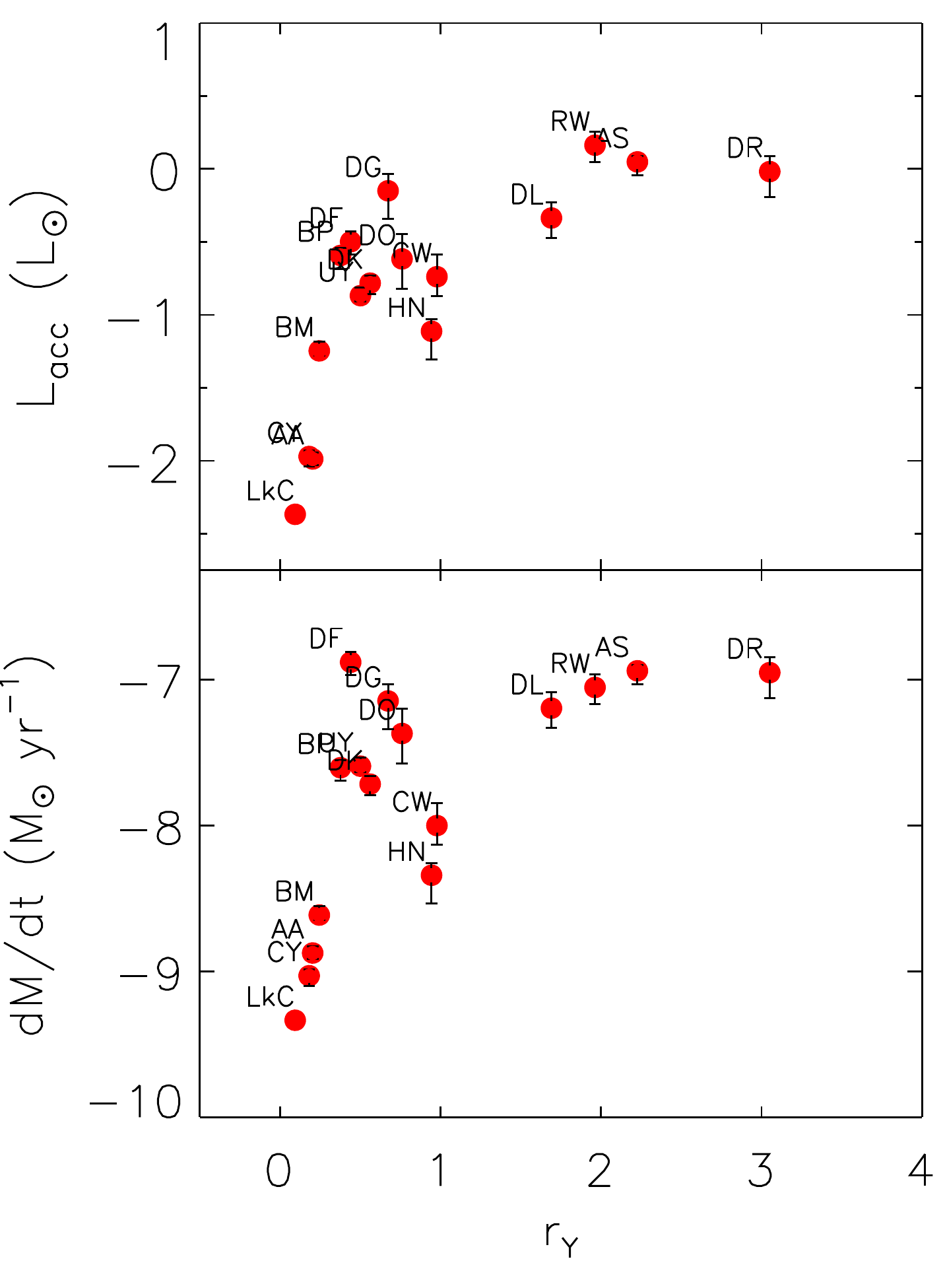}
\caption{Relation of the accretion luminosity (upper) and mass accretion rate (lower), as evaluated from published correlations with {\pbeta} luminosity, to the Y band veiling, $r_Y$, from SpeX spectra of 16 stars. Data points correspond to values based on the mean $A_V$ from the literature and error bars show how the quantities change over the full spread of $A_V$ reported for each star.  \label{f.acc}}
\end{figure}

Our SpeX spectra were not calibrated on an absolute scale, but the shape of the continuum is well defined by the relative fluxes and we used the SpeX continuum flux adjacent to each line plus an estimate for extinction to restore equivalent width ratios to the line intensity ratios that are the basis for comparing to models. The extinction is a much larger source of uncertainty. Table~\ref{t.Av} illustrates the magnitude of this problem. The $A_V$ in column 2 is from \citealt{fis11}, found from fitting the observed spectral energy distribution from 0.8 to 2.4~$\micron$ with a combination of a main sequence spectral template with zero reddening plus a continuum excess anchored by simultaneously measured line veilings of photospheric features. This technique, following the approach taken in \citet{gul98} for optical spectra, is in principle superior to deriving extinction from observed colors, since, as shown in \citealt{fis11}, most T Tauri stars have excess emission at all wavelengths, invalidating the standard technique of deriving $A_V$ from colors. However, as shown in \citealt{fis11}, when applied in the near infrared, this technique yields $A_V$ that are often larger than those derived from optical studies, due at least in part to the use of main sequence spectral templates and the possible presence of large cool spots.   Thus we also list in Table~\ref{t.Av}  the minimum and maximum $A_V$ collected from the literature, the mean of these values (including FEHK), and a new estimate of $A_V$ that we will derive in this paper in Section 5.3, based on comparing observed line ratios to the local line excitation models of \citet{kwan11}. The sources for the literature values of $A_V$ are cited in the notes to Table~\ref{t.Av}.  We use the extinction law of \citet{fit99} with $R_V=3.1$, as represented in the routine {\em fm\_unred.pro} in the IDL Astronomy Library, to convert between extinctions at different wavelengths.\footnote{http://idlastro.gsfc.nasa.gov/} 

Also listed in Table~\ref{t.Av}  are the accretion luminosities and mass accretion rates derived from the luminosity of {\pbeta}.  The {\pbeta} luminosities are derived from the line equivalent width, the absolute flux in the adjacent continuum, the extinction and the distance. Since we did not have absolute flux calibrated spectra we adopted continuum fluxes from the 2MASS $J$ magnitudes and used distances of 140 pc, 200 pc and 440 pc for Tau-Aur \citep{bertout06}, AS 353A \citep{rice} and BM And \citep{ave69} respectively.    The accretion luminosity and mass accretion rates were calculated from {\pbeta} luminosities following the calibration of \citet{muz98c} and \citet{natta04}. Mass accretion rates also required estimates of the stellar mass and radii, made from application of the \citet{siess} tracks to the effective temperatures and stellar luminosities from \citet{heg}. For extinctions we used the mean value of $A_V$ identified in Table~\ref{t.Av}. Both accretion luminosity and mass accretion rate are plotted in Figure~\ref{f.acc} against the simultaneously observed 1 $\mu$m veiling, $r_Y$. The symbols correspond to the values derived with the mean $A_V$ while the vertical lines extending below/above each symbol correspond to values derived from the minimum/maximum values of  $A_V$ from the literature cited in Table~\ref{t.Av}. The figure illustrates that even with the large spread in reported $A_V$, $r_Y$ is a reasonable proxy for accretion luminosity and disk accretion rate, where the accretion luminosities for our sample range from 1.5~\lsun\ to $4 \times 10^{-3}$ \lsun\ and the disk accretion rates from $2.3 \times10^{-10}$ to $1.3 \times 10^{-7}$  \msunyr .

We will adopt the $A_V$ from FEHK in calculating emission intensity ratios since it is determined in the same way for all stars in our sample. A difference of 1 magnitude in $A_V$ affects the ratios {\pgamma}$/${\pbeta} and {\bten}$/${\bgamma} by a factor of 7\%, {\bgamma}$/${\pbeta} by 12\%, and Pa12$/${\pbeta} by 19\%. There is no correlation between line ratios and $A_V$, indicating there is not a systematic effect on the ratios from the adopted extinction. An examination of the effect of extinction in the comparison of observed line ratios to those predicted from line excitation models will be the focus of Section 5.3.

\section{The Hydrogen Line Ratios}

In this section we present line intensity ratios for the Paschen series {\pbeta} through Pa12, and for two Brackett lines, {\bgamma} and {\bten}. We will compare the 1~$\mu$m veiling, $r_Y$, to observed line ratios and make use of the fact that the broad Paschen and Brackett lines are resolved with the modest SpeX velocity resolution of about 150 \kms.  

\begin{figure*}
\includegraphics[angle=90,width=\hsize] {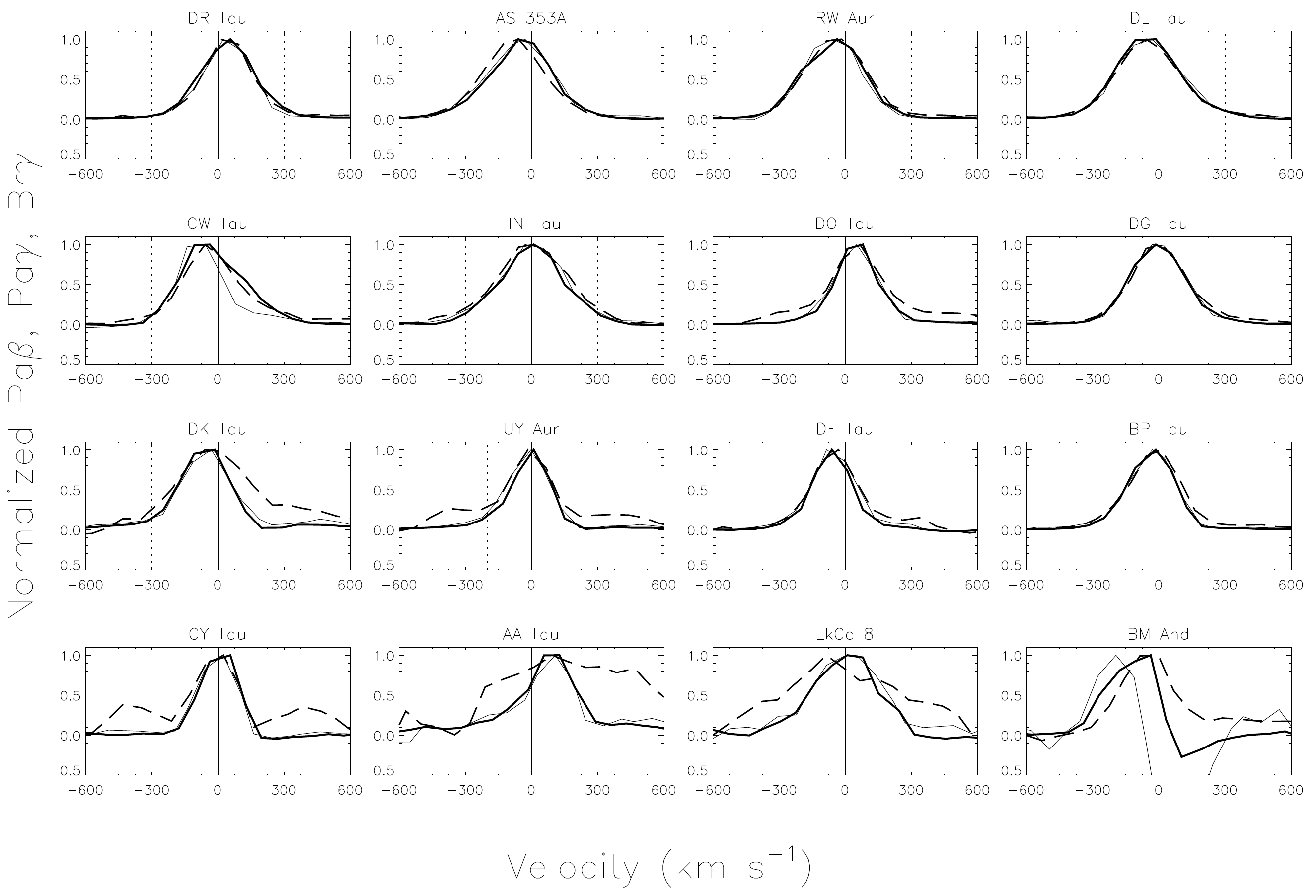}
\caption{Superposed normalized residual line profiles arranged in order of Y band veiling, $r_Y$, for {\pbeta} (dark solid line), {\pgamma} (light solid line)  and  {\bgamma} (dashed line).  Dotted vertical lines show the velocity limits selected for reliable emission line ratios, over velocity ranges where opacity differences will not confuse line ratios and the solid vertical line marks the stellar photospheric velocity. The weak and poorly defined {\bgamma} profiles of AA Tau, LkCa 8 and BM And led us not to compare them with models. \label{f.vel}}
\end{figure*}

We calculate intensity ratios for {\pgamma/\pbeta} and for {\bgamma/\pbeta} in three different ways for each star in our sample. One (`direct') is the intensity ratio based on the emission equivalent width measured directly from the SpeX spectra over the full range of the emission. The other two are measured from {\it residual line profiles}, where the photosphere of an appropriately veiled spectral template is subtracted from the CTTS spectrum (see \citealt{edw94} for a description of this technique and FEHK for the templates used here).  In one case the equivalent widths are measured over the full range of the residual emission, and in the other the equivalent widths, and corresponding line ratios, are found over a limited velocity range in the residual line profile. The line ratios from each of these three techniques are tabulated in Table~\ref{t.intensity} for each star. 

The residual {\pbeta}, {\pgamma}, and  {\bgamma} profiles (normalized and superposed) are shown in Figure~\ref{f.vel} with stars arranged in order of $r_Y$, from highest to lowest. Although the SpeX resolution of 150 {\kms} just resolves these broad lines, it is sufficient to show the opacity dependent differences in the redshifted absorption in some stars (e.g.\ CW Tau, DK Tau, DF Tau, BM And).   We therefore compute the intensity ratios from residual profiles both over the full range of emission (labeled `full') and also over velocity intervals selected to be free of opacity effects from the redshifted absorption (labeled `select'), in principle giving more reliable ratios when comparing to models. The velocity intervals are marked in Figure~\ref{f.vel} with vertical dotted lines and identified in Table~\ref{t.intensity}. Intensity ratios determined from emission equivalent widths in these velocity intervals are given for {\pgamma}$/${\pbeta}, {\bgamma}$/${\pbeta} and {\bten}/{\bgamma} in Table~\ref{t.intensity}  and the Paschen series through Pa12, relative to {\pbeta}, in Table~\ref{t.decrement}. We omit higher series lines because the lines are weak in most stars and the \citet{kwan11} calculations used a 15-level H atom, giving reliable fluxes only through Pa13 and Br13.

\begin{deluxetable*}{lcccccclll}
\tablecaption{Intensity Ratios and Selected Velocity Limits \label{t.intensity}}
\tabletypesize{\footnotesize}
\tablewidth{\hsize}
\tablehead{\colhead{Object}  & \colhead{ \pgamma/\pbeta} & \colhead{ \bgamma/\pbeta} & \colhead{ \pgamma/\pbeta} & \colhead{ \bgamma/\pbeta}  & \colhead{V Min} & \colhead{V Max} & \colhead{ \pgamma/\pbeta} & \colhead{ \bgamma/\pbeta} &  \colhead{Br10/\bgamma}  \\ 
\colhead{} &  \colhead{DIRECT} & \colhead{DIRECT} &\colhead{FULL} & \colhead{FULL} & \colhead{\kms} & \colhead{\kms} & \colhead{SELECT} & \colhead{SELECT}  & \colhead{SELECT} \\
\colhead{(1)} & \colhead{(2)} & \colhead{(3)} & \colhead{(4)} & \colhead{(5)} & \colhead{(6)} & \colhead{(7)} & \colhead{(8)} & \colhead{(9)} & \colhead{(10)} }
\startdata
AA Tau		&	2.23	&	0.55:	&	0.86	&	0.49:	&	0	&	150	&	0.81	&	0.22:	&	\nodata	\\
AS 353A		&	0.47	&	0.69	&	0.69	&	0.45	&	-400	&	200	&	0.68	&	0.27	&	0.82	\\
BM And		&	0.16	&	2.20:	&	0.21	&	3.25:	&	-300	&	-200	&	0.46	&	0.17:	&	\nodata	\\
BP Tau		&	0.57	&	0.36	&	0.64	&	0.32	&	-200	&	200	&	0.62	&	0.17	&	0.40	\\
CW Tau		&	0.44	&	0.71	&	0.61	&	0.70	&	-300	&	0	&	0.76	&	0.34	&	0.80	\\
CY Tau		&	1.11	&	0.39	&	0.69	&	0.33	&	-150	&	150	&	0.68	&	0.12	&	\nodata	\\
DF Tau		&	0.66	&	0.47	&	0.76	&	0.35	&	-150	&	0	&	0.66	&	0.19	&	0.57	\\
DG Tau		&	0.36	&	0.93	&	0.78	&	0.45	&	-200	&	200	&	0.76	&	0.25	&	0.68	\\
DK Tau		&	0.68	&	0.51	&	0.83	&	0.33	&	-300	&	0	&	0.73	&	0.18	&	\nodata	\\
DL Tau		&	0.49	&	0.59	&	0.81	&	0.34	&	-400	&	300	&	0.85	&	0.22	&	0.77	\\
DO Tau		&	0.45	&	0.61	&	0.68	&	0.36	&	-150	&	150	&	0.67	&	0.16	&	\nodata	\\
DR Tau		&	0.63	&	0.44	&	0.80	&	0.35	&	-300	&	300	&	0.80	&	0.24	&	0.70	\\
HN Tau		&	0.39	&	0.85	&	0.70	&	0.41	&	-200	&	200	&	0.68	&	0.22	&	0.37	\\
LkCa 8		&	2.64	&	0.54:	&	1.01	&	0.51:	&	-150	&	0	&	0.86	&	0.37:	&	\nodata	\\
RW Aur		&	0.46	&	0.50	&	0.71	&	0.36	&	-300	&	300	&	0.70	&	0.22	&	0.39	\\
UY Aur		&	0.58	&	0.41	&	0.69	&	0.35	&	-200	&	-200	&	0.64	&	0.16	&	0.19	
\enddata
\tablecomments{Cols.~2, 3:\ Intensity ratios based on directly measured equivalent widths, Cols.~4, 5:\ Intensity ratios based on {\it residual} emission profiles over the full velocity range of the emission, Cols.~6, 7:\ Selected velocity intervals for measuring line ratios to eliminate effects of opacity,  Cols.~8--10:\ Intensity ratios based on {\it residual} emission profiles over selected velocity intervals. Three stars in Cols.~3, 5, and 9 have errors in excess of 50\%; these are marked with a `:'.}
\end{deluxetable*}

\begin{deluxetable*}{lccccccc}
\tablecaption{Paschen Decrement from Selected  Velocity Limits \label{t.decrement}}
\tabletypesize{\footnotesize}
\tablewidth{\hsize}
\tablehead{
\colhead{Star} & \colhead{\pgamma/\pbeta} & \colhead{Pa7/\pbeta} & \colhead{Pa8/\pbeta} & \colhead{Pa9/\pbeta} & \colhead{Pa10/\pbeta} & \colhead{Pa11/\pbeta} & \colhead{Pa12/\pbeta}  \\
\colhead{} & \colhead{(2)} & \colhead{(3)} & \colhead{(4)} & \colhead{(5)} & \colhead{(6)} & \colhead{(7)} & \colhead{(8)}   } 
\startdata
AS 353A	&	0.68	&	0.62	&	0.54	&	0.55	&	0.51	&	0.42	&	0.41	\\
BP Tau	&	0.62	&	0.44	&	0.39	&	0.34	&	0.22	&	0.14	&	0.13	\\
CW Tau	&	0.76	&	0.63	&	0.57	&	0.63	&	0.51	&	0.45	&	0.41	\\
DF Tau	&	0.66	&	0.45	&	0.48	&	0.31	&	0.07	&	0.12	&	0.12	\\
DG Tau	&	0.76	&	0.67	&	0.64	&	0.55	&	0.48	&	0.46	&	0.37	\\
DK Tau	&	0.73	&	0.49	&	0.50	&	0.47	&	0.16	&	0.13	&	0.11	\\
DL Tau	&	0.85	&	0.72	&	0.72	&	0.74	&	0.64	&	0.55	&	0.46	\\
DO Tau	&	0.67	&	0.48	&	0.45	&	0.34	&	0.30	&	0.11	&	0.10	\\
DR Tau	&	0.80	&	0.71	&	0.59	&	0.54	&	0.48	&	0.42	&	0.35	\\
HN Tau	&	0.68	&	0.59	&	0.46	&	0.36	&	0.31	&	0.23	&	0.27	\\
RW Aur	&	0.70	&	0.64	&	0.48	&	0.24	&	0.34	&	0.27	&	0.28	\\
UY Aur	&	0.64	&	0.49	&	0.42	&	0.22	&	0.22	&	0.17	&	0.11	
\enddata
\tablecomments{Paschen line ratios relative to {\pbeta} for 13/16 stars from our sample measured over the velocity intervals identified in Table 3. Higher Paschen lines are weak or not detected in AA Tau, BM And and LkCa 8.} 
\end{deluxetable*}	

\begin{figure}
\includegraphics[angle=90,width=\hsize] {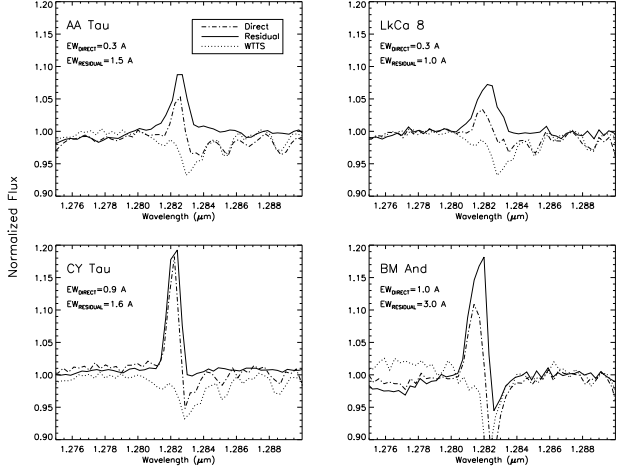}
\caption{Comparison of {\pbeta} for direct (solid) and residual (dash-dot) profiles, along with the WTTS template (dotted) for the four stars with smallest {\pbeta} equivalent width in the direct spectra. In these four stars the underlying photospheric absorption affects the observed profile and the residual profiles have different equivalent width and kinematic structure. For the remaining 12 stars the difference between direct and residual profiles is $\le$ 10\%.  \label{f.residual}}
\end{figure}

For most of our sample the difference between direct and residual profiles is $<$10\%. However for the four stars with the smallest emission equivalent width and lowest $Y$-band veiling ($r_Y \le 0.2$, AA Tau, BM And, CY Tau and LkCa 8), these two approaches yield significant differences in profile morphology and equivalent width, as illustrated in Figure~\ref{f.residual} for {\pbeta}. Three of these four stars show residual {\bgamma} profiles in Figure~\ref{f.vel} that are poorly defined, and their small line to continuum ratios and equivalent widths yield equivalent width errors in the selected velocity intervals $\ge$ 50\%. We flag these large errors in the Table and  exclude these three stars in the subsequent figures that include the ratio of {\bgamma}$/${\pbeta}. One other star,  CY Tau, also shows a noisy residual {\bgamma} profile with a small {\bgamma} equivalent width in the selected velocity interval with an error around 30\%.  Although we include this star in the figures, the error in its {\bgamma}$/${\pbeta} ratio is significant in comparison to the other stars, and as will be seen in later sections it is an outlier in some relationships. We note that in contrast to previous papers that focus only on {\pbeta} and {\bgamma}, here we put {\pbeta} in the denominator of our ratios rather than {\bgamma}. This facilitates comparisons among a wider range of lines, and is also more meaningful in comparison to the model predictions discussed in the next section. 

\begin{figure}
\includegraphics[angle=90,width=\hsize] {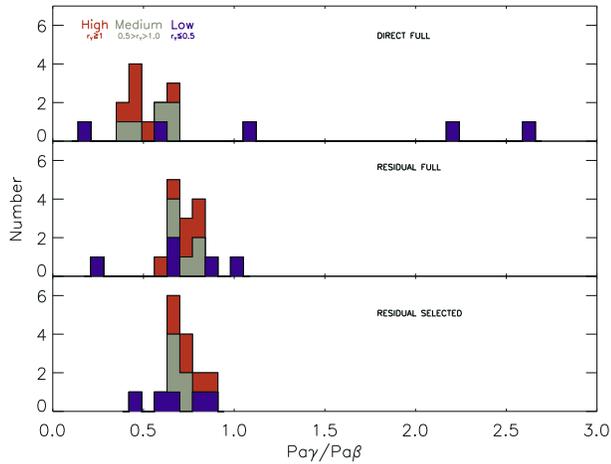}
\caption{Distribution of {\pgamma}$/${\pbeta} line intensity ratios for the 16 stars in our SpeX sample based on three assessments of the emission equivalent width (EW). Upper panel:\ EW measured directly from the original spectrum over the full range of velocity. Middle panel:\ EW measured from the residual profile over the full range range of velocity. Bottom panel:\ EW measured from the residual profile, over selected velocity intervals shown in Figure 2. The spread in ratios and the mean value are reduced when selected velocity intervals from the residual profiles are used. The colors correspond to three levels of $Y$-band veiling; high is red with $r_Y \ge 1$, medium is gray with $ 0.5 \le r_Y < 1 $  and low is blue:\ $r_Y < 0.5$.  \label{f.hist}}
\end{figure}

We show in Figure~\ref{f.hist} a comparison of the {\pgamma}$/${\pbeta} emission intensity ratios for the 16 sample stars from the three approaches (direct full, residual full, residual selected)  in histogram form, where veiling groups identified as high ($r_Y \ge1$), medium ($0.5 \le r_Y <1$), and low ($r_Y < 0.5$) are separately colored. The figure shows that the dispersion in the intensity ratio is largest for the `direct' method and smallest for the `selected' method, suggesting that the range of actual line ratios may be smaller than would be inferred from the standard approach.  We will adopt the narrower range of ratios from selected velocity intervals in residual profiles for the analysis in the next section in the interest of making the best comparison to model predictions.

\begin{figure}
\includegraphics[width=\hsize] {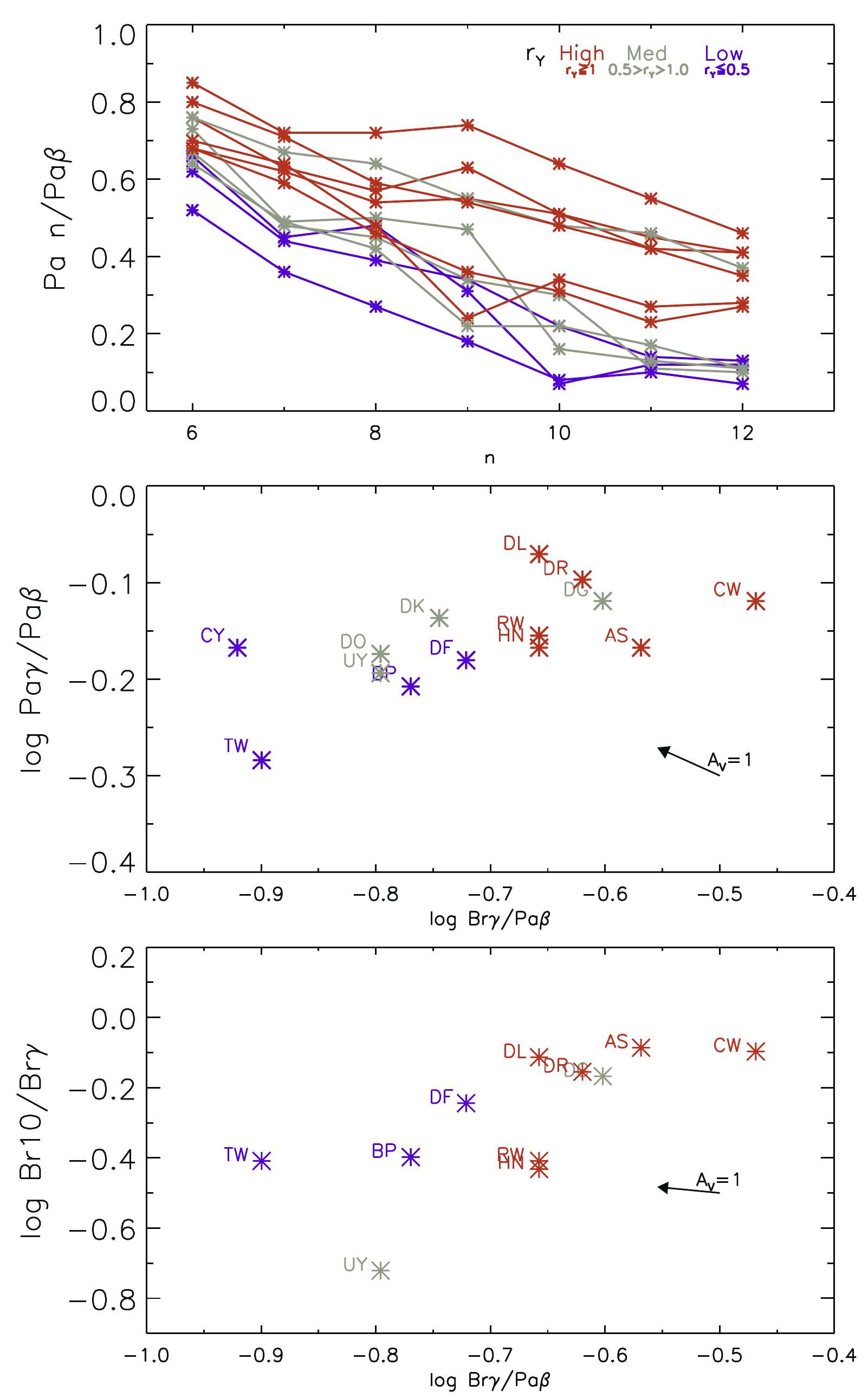}
\caption{Observed relations for three diagnostics.  Upper:\ Paschen decrement (14 stars), Middle:\ Ratio-ratio relation for {\pgamma}$/${\pbeta} versus {\bgamma}$/${\pbeta} (14 stars), Lower:\ Ratio-ratio relation for  {\bten}$/${\bgamma} versus {\bgamma}$/${\pbeta} (11 stars).  Color indicates levels of $Y$-band veiling:\ red for $r_Y \ge 1$, gray for $ 0.5 \le r_Y < 1 $  and blue for $r_Y < 0.5$. There is a tendency for stars with higher $r_Y$ to have larger line ratios. \label{f.relations}}
\end{figure}

We identify three relations in Figure~\ref{f.relations} that will be the basis for comparing observations to the theoretical predictions from both Case B and the \citet{kwan11}, local line excitation calculations. These are the Paschen decrement normalized to {\pbeta}, for {\pgamma} through Pa12, plus two ratio-ratio relations, one between {\pgamma}$/${\pbeta} and {\bgamma}$/${\pbeta}, and the other between  {\bten}$/${\bgamma} and  {\bgamma}$/${\pbeta}, again sorted into high, medium and low $Y$-band veiling groups.  

Because three low veiling objects in our sample of 16 CTTS have unreliable {\bgamma}$/${\pbeta} ratios, four have unreliable Paschen decrements, and six have unreliable {\bten}$/${\bgamma} ratios we introduce in this and all subsequent figures SpeX-based line ratios for one additional low veiling CTTS taken from the literature. This is TW Hya, an object with $A_V=0$, using line intensity ratios taken from \citet{vac}. We have a number of $Y$-band NIRSPEC spectra for this object taken between 2006 to 2011 and consistently find the veiling to be very low or zero \citep{edw06}, and adding it improves the statistics for the low veiling group. (Although our NIRSPEC profiles of TW Hya show a weak redshifted absorption at {\pgamma}, the ratios from Vacca and Sandell correspond to  the `direct full' ratio method defined here.)  

We see in all three relations that there is a spread in the observed ratios with a tendency for stars with higher $r_Y$ to have higher ratios than stars with lower $r_Y$.   However, with such a small sample it is not clear how robust this result is. We note that two of the low veiling stars from our sample that do not appear in  Figure~\ref{f.relations} due to their highly uncertain {\bgamma}$/${\pbeta} ratios (AA Tau and LkCa 8) have {\pgamma}$/${\pbeta} ratios at the high end of the observed values, suggesting that the trend we see here with veiling needs to be tested further. We also note that the slight rise in the Paschen decrement at Pa9 is an artifact due to blending from adjacent emission lines in the stars with higher veiling and stronger line emission. 

\section{Comparison of Case B and Local Line Excitation Models}

Hydrogen line ratios depend on physical conditions in the line formation region.  Baker and Menzel's Case B for radiative ionization and recombination has been successfully applied to understanding ionized nebulae such as HII regions and planetary nebulae for many decades and has recently been invoked to infer physical conditions in T Tauri systems from line ratios of the Paschen and Brackett series of hydrogen (see Introduction), making use of the interactive online server that calculates hydrogen line ratios for a range of electron densities and temperatures \citep{B}.  The results, primarily focusing on the behavior of series decrements, suggest a surprisingly diverse range of implied electron densities and temperatures in the hydrogen line formation region that calls into question the applicability of the Case B assumptions for T Tauri stars. The recent local line excitation calculations of \citet{kwan11}, hereafter KF, developed to interpret T Tauri spectral lines,  offer an alternative option for inferring physical conditions from observed hydrogen line ratios. Here we review both approaches, and in the next section compare predicted ratios to observed values in both scenarios.

The line diagnostics we will explore in each scenario arise from a similar range of electron density and temperature, however the physical conditions are actually quite different. The Case B calculations output line ratios as a function of input electron density $n_e$ and temperature $T$. No restrictions on $n_e$ and $T$ are explicitly imposed by the Case B online simulator \citep{B}. By definition, however, the recombination model needs the population of the $n=2$ level to be sufficiently small that collisional excitation from that level as a means  of Balmer, Paschen, Brackett, etc.\ line emission is not significant.   This stipulates that the neutral hydrogen column density, $n_{HI} \delta l$ (where $\delta l$ is the emission length scale), or equivalently, the Ly$\alpha$ optical depth, be sufficiently small that radiative de-excitation of $n=2$ occurs more rapidly than collisional excitation from $n=2$.  Since in Case B $n_e$ is a free parameter, the ratio $n_{H I} /n_e$  is not specified or determined,  nor is the photoionization rate from the ground state, $\gamma_{HI}$. Thus the recombination model implicitly assumes that $\gamma_{HI}$ and $\delta l$ have values that ensure sufficiently small line optical depths. 

The KF calculations are more general in including both recombination and collisional excitation as a means of producing line photons and in exploring the full range of line optical depths.  Atomic parameters include 15 distinct energy levels of hydrogen, 19 of \ion{He}{1} and key transitions of \ion{Ca}{2}, \ion{O}{1} and \ion{Na}{1}.  Upon inputs of the local physical conditions of hydrogen nucleon number density $n_H$, temperature $T$, ionization rate $\gamma_{HI}$ (photoionization rates from excited states are also included based on an assumed stellar plus veiling continuum), and the velocity gradient $dv/dl$ (giving the emission length scale in a differentially  moving medium),  the calculations  solve for the ionization fraction $n_e /n_H$, level population (including $n=1$), and all line optical depths self-consistently. The full set of input parameters, with $T$ from 5000 to 30,000 K, $n_H$ from $10^8$ to $2 \times 10^{12}$ cm$^{-3}$, $\gamma_{HI}$ of $2 \times 10^{-4}$ or $2 \times 10^{-5}$ $s^{-1}$, and $dv/dl$ of 150 {\kms}/2$R_*$ or 150 {\kms}/1.25 $R_*$, presented in KF are designed to approximate conditions in the region of a wind or an accretion flow where the bulk velocity is $\sim$ 150 {\kms}. The variations of the line emissivity ratios arising from the choices of $\gamma_{HI}$ and $dv/dl$ are very much  smaller  than those  arising  from density  and temperature changes, and the results shown here are from the case of $\gamma_{HI}$ = $2 \times 10^{-4}$~s$^{-1}$ and $dv/dl$ = 150 {\kms}/2$R_*$. Although each line emissivity is for a uniform density and temperature in roughly the middle of an accretion flow/wind, while the observed line flux will be an integration of the emissivity over the entire kinematic structure, KF pointed out that since each position in the flow will be represented by the local line excitation, only with different parameters, that it is possible to judge how the resultant ratios will be affected when averaged over a range of density and/or temperature. They concluded that observed line flux ratios do indeed indicate clearly enough the physical conditions and that this approximation is reasonable.

\begin{figure}
\includegraphics[width=\hsize] {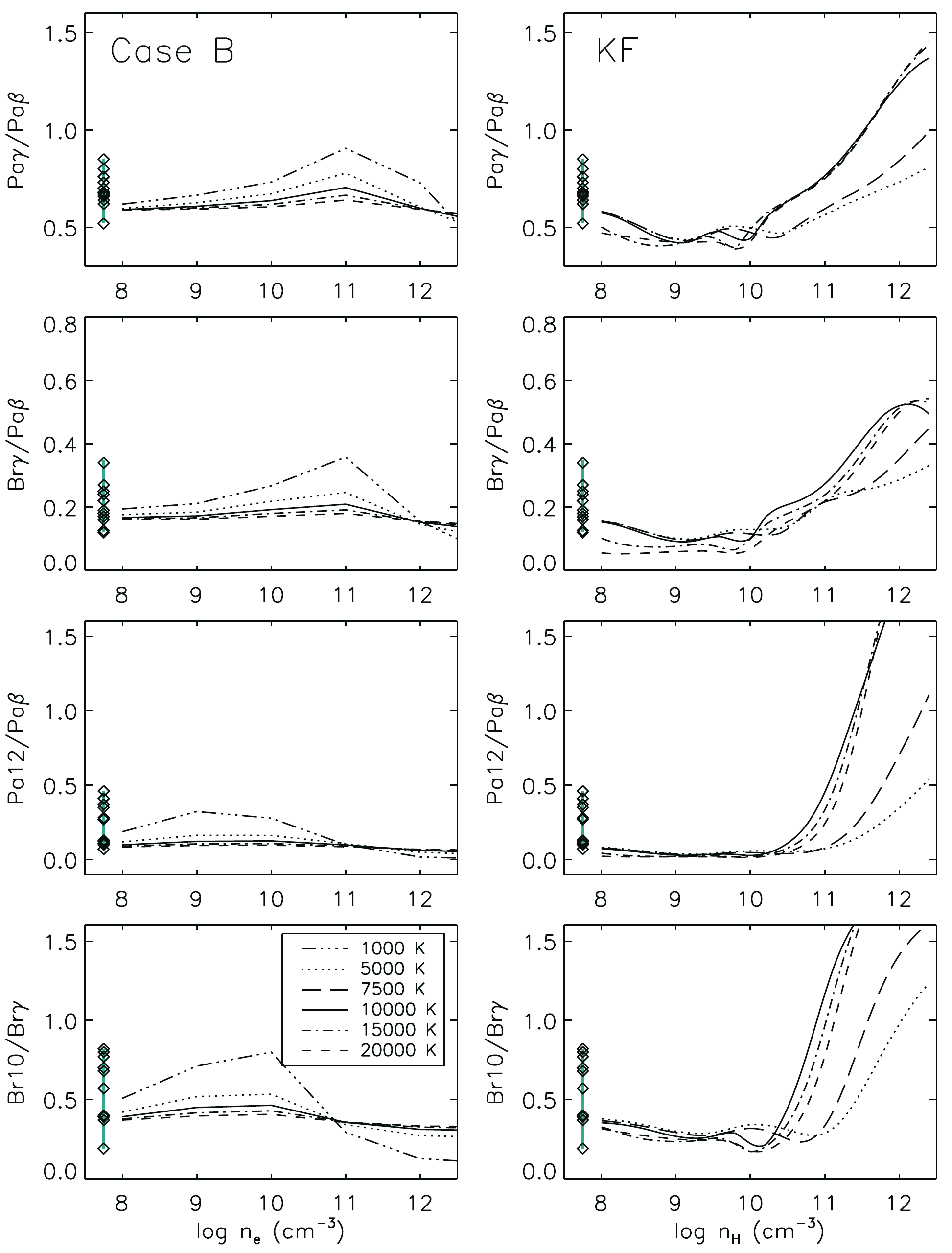}
\caption{The behavior of various line ratios as a function of $n_e$ for Case B (left, $T=1000$ -- 20,000 K) and  $n_H$ for KF  (right, $T=5000$ -- 20,000 K). In Case B, the line ratios have a very small dynamic range for $T<1000$ K.  In the KF calculations, the line ratios are initially low when the lines are optically thin and then increase steeply as $n_H$ increases, until they plateau at high optical depths. The range of observed ratios is shown along the left side of each panel. Temperature line types apply to all panels. \label{f.theory}}
\end{figure}

The two models have similar ranges in $n_e$, but refer to quite different physical regimes, arising from the fundamental difference in the energy source for the line photons, which is continuum photons more energetic than 13.6 eV  in the recombination model and thermal kinetic energy in the KF model. The derived electron fraction in the KF calculations is somewhat dependent on the ionization rate, as shown in their Figure 5,  but $n_e/n_H $ is $\ge0.6$ over a wide range of density for $T \ge$ 8750 K and $\sim$ 0.1 at $T = 7500$~K.  Thus although $n_e$  in both the Case B and KF models is similar,  the line optical depths are vastly different.  In KF, at fixed values of  $\gamma_{HI}$, $dv/dl$, and $T$, the run of the calculation with increasing $n_H$ corresponds to increasing line optical depths, producing a corresponding variation of the line emissivities.  
 For example, the Ly$\alpha$  optical depth for the case of $n_H=10^{11}$ cm$^{-3}$ and $T=$ 10,000  K   is $2\times  10^7$.  This  is much  higher  than the limit  imposed by the recombination model, since in KF collisional excitation from $n=2$, whose population is sustained by the strong Ly$\alpha$  trapping,  is the predominant  cause for the strength  of the line emission.   Consequently, the behavior of hydrogen line ratios with density and temperature is quite different in the two models. This is illustrated in Figure~\ref{f.theory} where the relation between increasing $n_e$ for Case B or $n_H$ for the KF local line excitation is shown for four ratios:\ {\pgamma}$/${\pbeta}, {\bgamma}$/${\pbeta}, Pa12$/${\pbeta}, and Br10$/${\bgamma} for a range of temperatures. For the KF models temperatures are shown from 5000 K to 20,000 K but for Case B we include temperatures as low as 1000 K in order to reach the full range of the observed ratios. The dispersion in observed values from Tables~\ref{t.intensity} and \ref{t.decrement} is indicated by a vertical line along the left side of each panel.  The observed ratios cover about a factor of two for {\pgamma}$/${\pbeta} and {\bgamma}$/${\pbeta} and about a factor of four for Pa12$/${\pbeta} and Br10$/${\bgamma}. 

For the KF models, the four ratios have the same general behavior with density and temperature, transitioning near $n_H \sim10^{10}$ cm$^{-3}$ from low optically thin values to higher ones, climbing steeply as the density and line optical depth increase. (See figures in KF for corresponding optical depths). The rise begins at lower densities for higher temperatures, but the general behavior is similar for temperatures between 5000 and 20,000 K. These relations readily account for the range of observed ratios, where higher ratios are explained by higher densities/optical depths. At the highest density in the KF calculations of $n_H = 2\times10^{12}$ cm$^{-3}$, the ratios of {\pgamma}$/${\pbeta} and Pa12$/${\pbeta} exceed unity and are still climbing, while the ratios of {\bgamma}$/${\pbeta} and Br10$/${\bgamma} have plateaued at peak values around 0.6 and 2, respectively.

In contrast, the Case B predictions show quite different behavior. At low $n_e$  all four ratios are higher when $T$ is lower.  This is because the radiative recombination rate to level $n$ is proportional to $1/T^{\alpha}$, where the index $\alpha$ increases from 0.7 to 1.1 as  $n$  increases  from 3 to 12, so as $T$  decreases  there  is a  stronger preference, comparatively, towards radiative recombinations into higher $n$ levels. As  $n_e$ increases, three-body recombinations begin to contribute.  At a given $T$ these recombinations, unlike radiative ones, strongly favor population into higher $n$ levels, since the rate to level $n$ is proportional to $n^2$  times a factor which itself increases rapidly with increasing $n$.  This rate is also roughly proportional to $1/T$, so its influence is also stronger when $T$ is lower. Hence the dispersion of each depicted line ratio with temperature increases with increasing $n_e$ until collisional de-excitation and collisional ionization begin to dominate over radiative decay and the ratio falls with further $n_e$  increase.  This transition occurs at lower $n_e$ for higher $n$ levels because collisional rates are higher while Einstein $A$ rates are lower for higher $n$.  While it occurs near $n_e \sim 10^{11}$ cm$^{-3}$ for {\pgamma}$/${\pbeta} and {\bgamma}$/${\pbeta}, in the case of Pa12$/${\pbeta} it occurs at $n_e \sim  10^9$  cm$^{-3}$.  (The density grid in the Storey and Hummer online server can interpolate ratios only between integer values of log $n_e$, so whether the peak ratio is precisely at the stated value is unclear).  
Note that the Case B relations require temperatures $\sim$ 1000 K  to account for ratios above the median observed value for each ratio shown, while electron densities $n_e\sim10^{11}$ cm$^{-3}$ are favored for  {\pgamma}$/${\pbeta}  and {\bgamma}$/${\pbeta} but $n_e$ $<$ $10^{10}$ cm$^{-3}$  is required for ratios incorporating the higher levels Pa12$/${\pbeta} and Br10$/${\bgamma}.

Thus, the differing behaviors of the line ratios with increasing $n_e$ shown in Figure~\ref{f.theory} result from  the fundamentally different nature of the two calculations.   In the KF calculations the behavior follows simply from the fact that as $n_H$ increases at a local point with all the other parameters fixed, the line optical depths also increase.  The rise of each ratio with $n_H$ follows from the stronger build-up of population into higher $n$ levels via collisions, coupled with the larger optical depth of the lower transition, thus enhancing the emissivity of the line in the numerator and reducing the emission effectiveness of the line in the denominator. 

In the recombination model, there is no explicit determination of $n_{HI}$, but given that $n_{HI}$$\gamma_{HI} = n_p (n_e\alpha_{\rm rad}+ n_e^2\alpha_{3b}$), where $\alpha_{\rm rad }$ and $\alpha_{3b }$ are the radiative and three-body recombination coefficient respectively, $n_p$ is the proton density and neglecting, for ease of explanation, the process of collisional ionization from high $n$ levels, $n_{HI}$ will, at fixed $\gamma_{HI}$,  increase rapidly  as $n_e$  increases,  so to maintain  the same  constraint  on $n_{HI} \delta l$ in order for recombination  to remain  dominant  over collisional excitation  from $n$ = 2 as a means of photon production, $\gamma_{HI}$ must increase or $ \delta l$ must decrease accordingly.  Indeed, with collisional rates  increasing  with  increasing  $n_e$  while Einstein  $A$ rates  remain  fixed, the limit on $n_{HI} \delta l$  actually needs to vary inversely with $n_e$, making the required changes in $\gamma_{HI}$ and/or $ \delta l$ even more drastic.  Thus  the recombination  results  actually implicate  associated changes in those two parameters which one must also assess for plausibility before applying the Case B results to observations.  For example, if the length scale of the emission corresponding to the same velocity gradient used in KF is adopted,  the resultant constraint on the Ly$\alpha$  optical depth of $\le 10^5$ requires $n_{HI}  \le 4 \times 10^7$ cm$^{-3}$ for the Case B conditions.   Then,  for $T \le$ 5000 K , $\gamma_{HI} \ge$ ($n_e/10^{10})^2$ s$^{-1}$. For $n_e = 10^{10}$ cm$^{-3}$, for example, the photoionization rate from the ground state must then be orders of magnitude higher than the $\gamma_{HI}$= $2 \times 10^{-4}$ s$^{-1}$ considered in the KF calculations. (The fairly high ionization in KF with  $n_e /n_H$  of $\ge$ 0.6 for $T \ge 8750$ K  is due to photoionization from excited states by the stellar and veiling continuum.)  The length scale of the emission region can be reduced  to alleviate the constraint  on $\gamma_{HI}$, but probably  not by more than a factor of 10. Thus in Case B it is the requirement that recombinations continue to predominate at high $n_e$, demanding a physical regime requiring particularly strong photoionizations to maintain low line optical depths, that is the reason for the depicted  behavior of the line ratios with $n_e$ in Figure~\ref{f.theory}.

These restrictions embedded within the Case B assumptions are grounds to doubt their validity in the interpretation of hydrogen line ratios in CTTS. To further demonstrate this point, in the next section we compare conclusions of both Case B and the KF calculations to hydrogen line ratios.

\section{Comparison of Observations to Case B and Local Line Excitation Models }

Here we return to the three diagnostics shown in Figure~\ref{f.relations}. For both Case B and the KF local line excitation calculations, we compare observational diagnostics to theoretical predictions for (1) the Paschen decrement normalized to {\pbeta}, for {\pgamma} through Pa12, (2) the ratio-ratio relation of {\pgamma}$/${\pbeta} and {\bgamma}$/${\pbeta}, and (3) the ratio-ratio relation of {\bten}$/${\bgamma} and  {\bgamma}$/${\pbeta}.  

\subsection{Case B}

\begin{figure}
\includegraphics[width=\hsize] {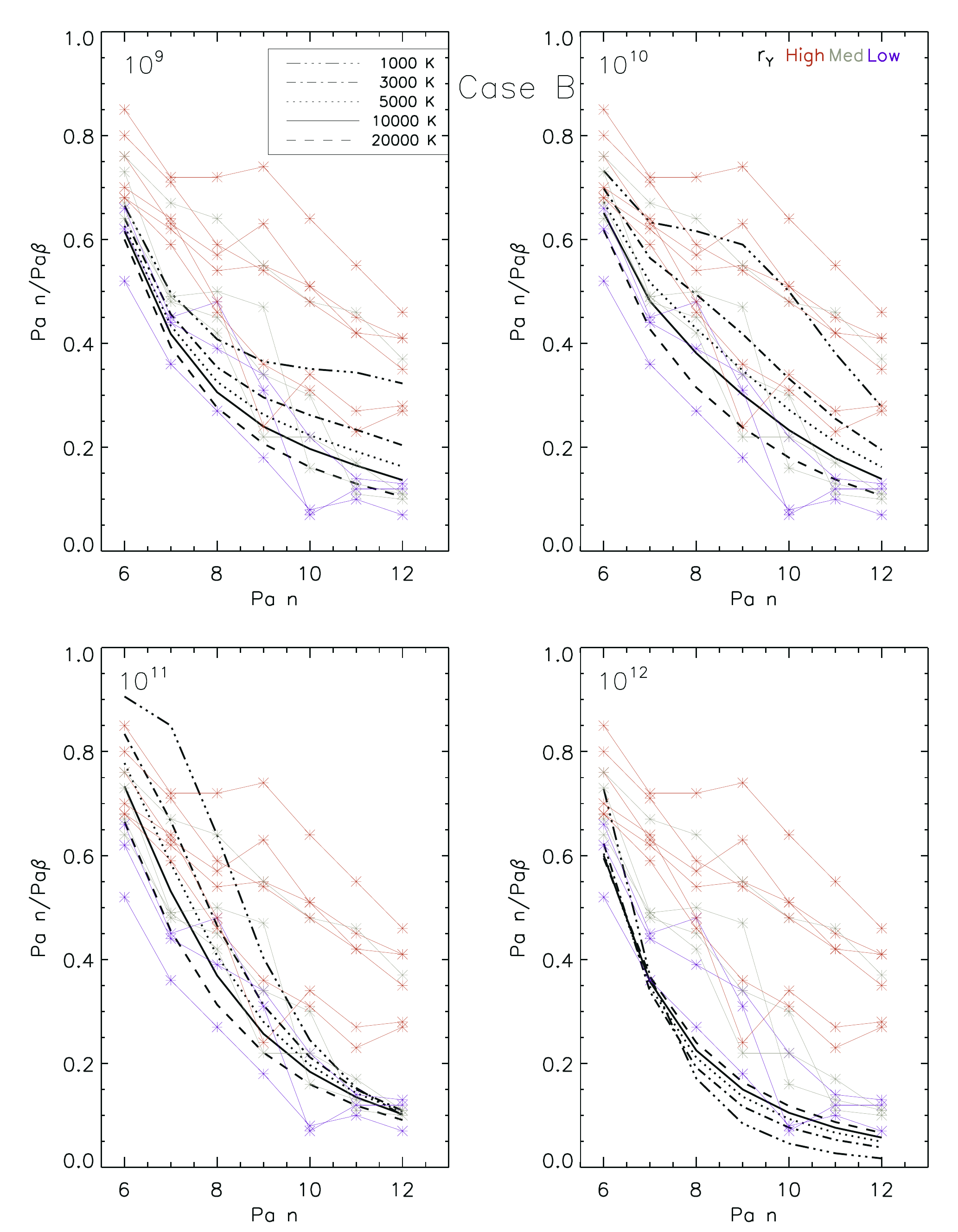}
\caption{Paschen decrement for Case B compared to observations of 12 stars (Table~\ref{t.decrement}). Each panel features a different log $n_e$ (9, 10, 11, 12), for a range of temperatures from 1000 to 20,000 K. \label{f.B_NH}}
\end{figure}

\begin{figure}
\includegraphics[width=\hsize] {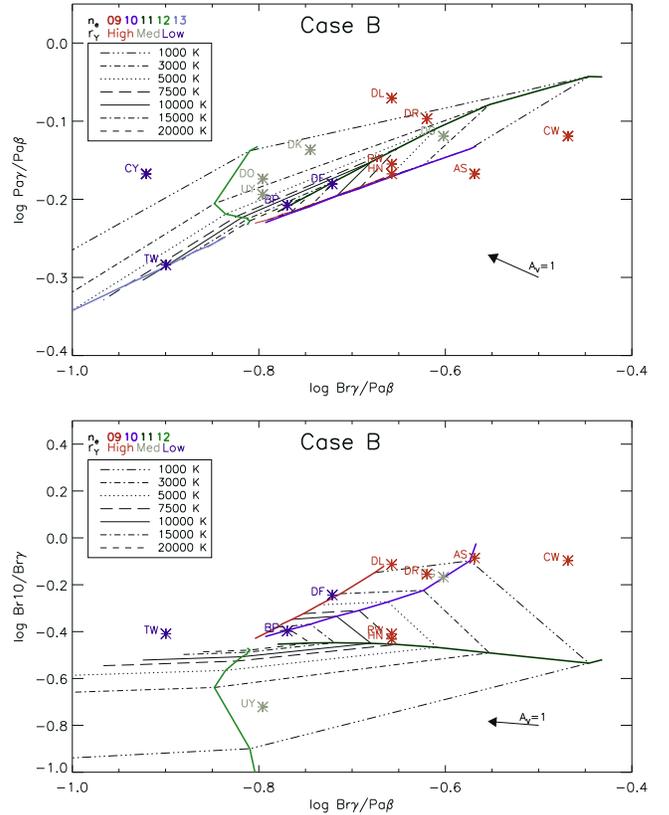}
\caption{Case B ratio-ratio relations for $n_e$ and $T$ compared to observations. {\it Upper}:\ Theoretical relations for {\pgamma}$/${\pbeta} versus {\bgamma}$/${\pbeta} are shown in different line types for temperatures between 1000 K and 20,000 K. Iso-density lines connect ratios for $\log n_e = 9$, 10, 11, 12, 13  ($\log n_e= 9$, 10 are degenerate).  {\it Lower}:\  Similar relations for {\bten}$/${\bgamma} versus {\bgamma}$/${\pbeta} ($\log n_e =13$ is out of the plot range). In both panels observations are colored according to $r_Y$ group. \label{f.B_ratios}}
\end{figure}

The comparison between Case B predictions and observations is shown Figure~\ref{f.B_NH} for the Paschen decrement and in Figure~\ref{f.B_ratios} for the two ratio-ratio relations. We have clarified in the previous section the rather stringent requirements that allow Case B to be applied as $n_e $ increases but continue with the comparison since this model has been widely applied to observed CTTS line ratios. In order to keep the figure from being too busy we restricted the range of $n_e$ (in units of cm$^{-3}$) to four orders of magnitude, adopting the range most often cited in comparison to observed ratios, for log $n_e = 9$, 10, 11, 12. The predicted Paschen decrement in Figure~\ref{f.B_NH} is shown in four panels, one for each electron density, each with iso-temperature surfaces from 1000 to 20,000 K.   The predicted decrements differ primarily in the maximum value of {\pgamma}$/${\pbeta} (highest for $\log n_e= 11$) and in the rate of decline from lower to higher Paschen lines (shallowest for  $\log n_e = 10$ and $T < 5000$ K). The higher veiling stars, with a combination of a high {\pgamma}$/${\pbeta}  ratio and a shallow decline down the series, are not well matched by Case B. The lower veiling stars are more readily accounted for, with small {\pgamma}$/${\pbeta} and a steep decrement to Pa12 for a range of electron densities and temperatures from 10,000 to 20,000 K. 

The iso-density lines behave differently in each ratio-ratio plot in Figure~\ref{f.B_ratios} due to the different behavior of line ratios in Case B with increasing $n_e$ seen in Figure~\ref{f.theory}. In the {\pgamma}$/${\pbeta} versus {\bgamma}$/${\pbeta} relation, the locus of iso-density lines rotates counter-clockwise around the figure, with the widest range of allowed ratios for the lowest temperatures ($T = 1000$ K). In contrast in the {\bten}$/${\bgamma} versus {\bgamma}$/${\pbeta} relation the locus of iso-density lines rotates clockwise around the figure, and again the widest range of allowed ratios is for the lowest temperatures ($T = 1000$ K). The observational trend for a pair of ratios to increase together is largely a temperature effect in Case B, where the highest ratios require temperatures below 3000 K. However, an uncertainty of $A_V$ of only one magnitude is sufficient to move an individual star through a wide range of  $n_e$ and temperature loci, putting stringent, and probably unrealistic, requirements on the precision of $A_V$ in order to make a reliable comparison. With that caveat, most observed points in the {\pgamma}$/${\pbeta} versus {\bgamma}$/${\pbeta} relation suggest log $n_e$  from 10--12  and temperatures over the full range from 1000 to 20,000 K.

These results can be compared with previous studies comparing line ratios to Case B. The conclusion that the best Case B fits to series decrements required $\log n_e = 10$  and $T = 3000$ K by \citet{bary08} was based on averaging line ratios for a group of Tau-Aur stars with high veiling. While this conclusion is obtained under Case B to match the shallow fall-off of the series decrement in the high veiling stars (but not the low veiling stars), the observed  {\pgamma}$/${\pbeta} is too high in comparison to the Case B predictions for these conditions.  The Case B fit for Paschen and Brackett series decrement found by \citet{vac} for the low $Y$-band veiling star TW Hya,  with $\log n_e  = 13$  and $T =$ 20,000 K, is consistent with its comparison with all three diagnostics here. However most CTTS would not be compatible with these findings for TW Hya. 

In sum, Case B predictions are questionable for CTTS.  The problem is most severe for the high veiling stars which are not well described by the predicted Paschen decrement while the ratio-ratio plots suggest a wide range of electron densities and temperatures among the stars. The low veiling stars are reasonably well matched to predicted Paschen decrements for $\log n_e = 11$ and $T>$ 10,000 K,  although in the ratio-ratio plots, except for TW Hya, they require electron densities at least an order of magnitude lower and a range of temperatures. This is consistent with the diverse range of temperatures and electron densities reported in the literature for classical T Tauri stars when being compared to Case B predictions.

\begin{figure}
\includegraphics[width=\hsize] {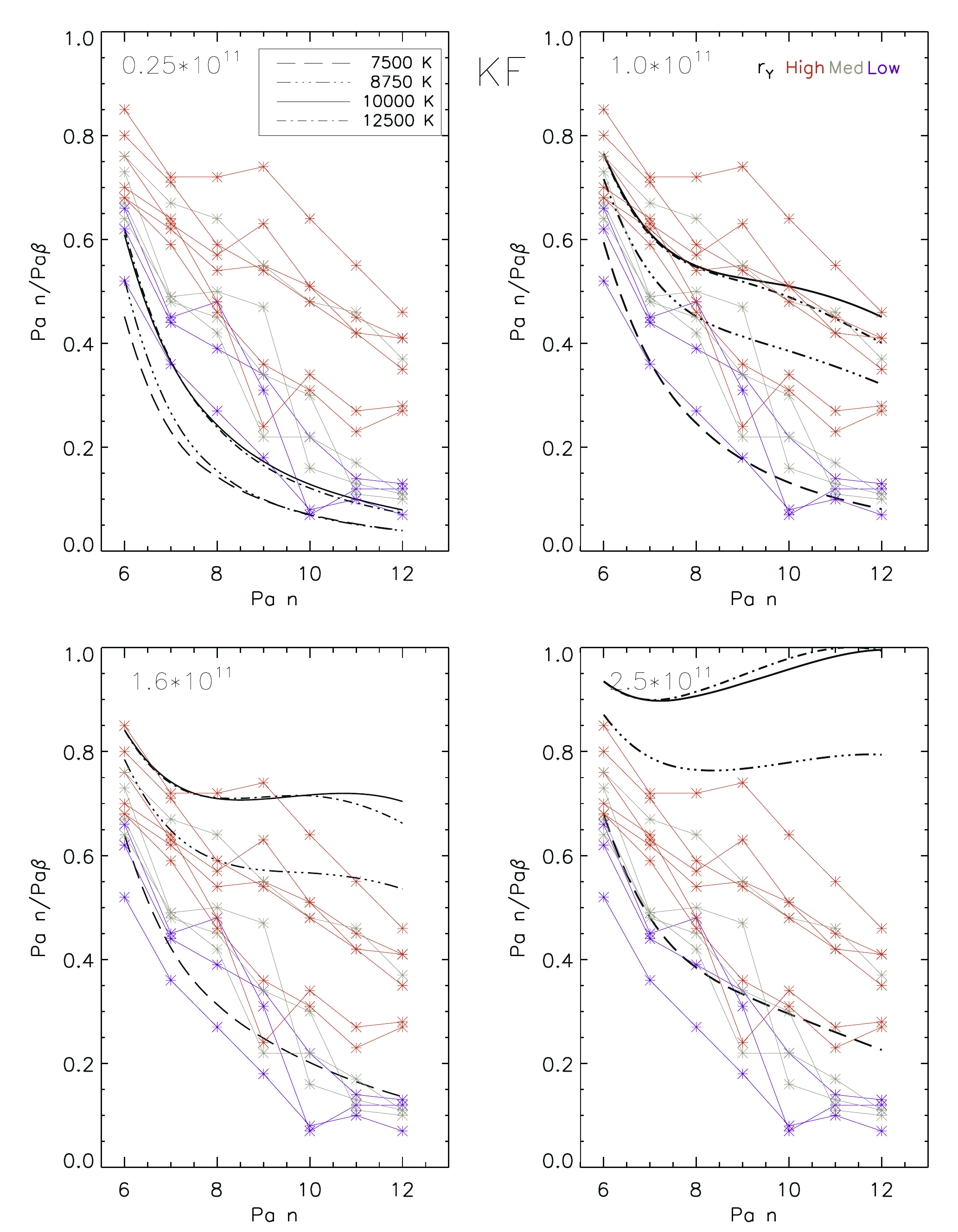}
\caption{Paschen decrement for KF compared to observations of 12 stars (Table~\ref{t.decrement}). Each panel features a different $n_H$ (0.25, 1, 1.6, 2.5 $\times 10^{11}$), for a range of temperatures from 7500 to 12,500~K.\label{f.KF_NH}}
\end{figure}

\begin{figure}
\includegraphics[width=\hsize] {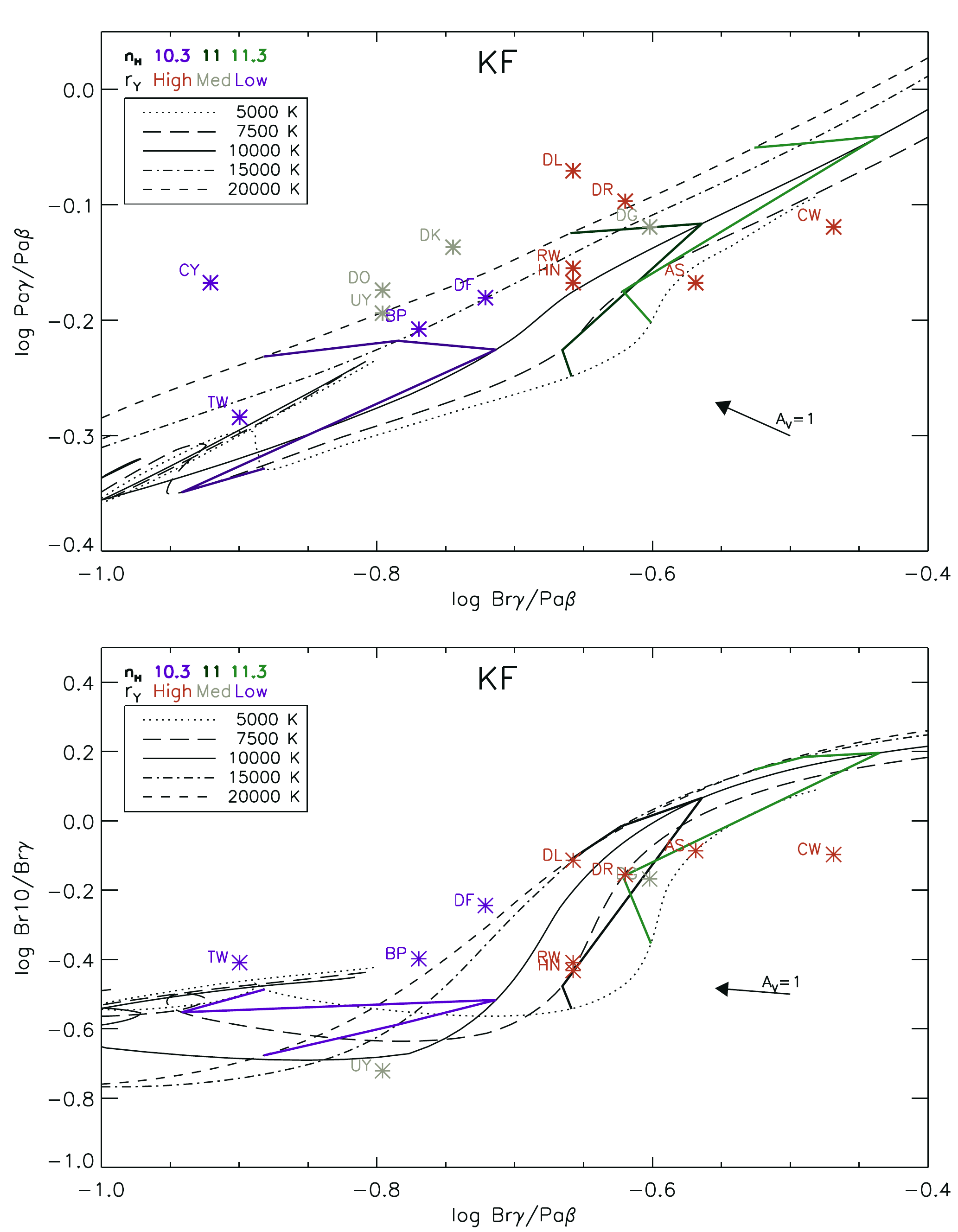}
\caption{KF ratio-ratio relations compared to observations. {\it Upper}:\ Theoretical relations for {\pgamma}$/${\pbeta} versus {\bgamma}$/${\pbeta} are shown in different line types for temperatures between 5000 K and 20,000~K. Iso-density lines connect ratios for $\log n_H = 10.3$, 11, 11.3.  {\it Lower}:\  Similar relations for {\bten}$/${\bgamma} versus {\bgamma}$/${\pbeta}. In both panels observations are colored according to $r_Y$ group. \label{f.KF_ratios}}
\end{figure}

\subsection{Local Line Excitation}

The comparison between the Kwan \& Fischer (KF) local line excitation predictions and observations is shown in  Figure~\ref{f.KF_NH} for the Paschen decrement and in Figure~\ref{f.KF_ratios} for the two ratio-ratio relations. Predicted decrements are again shown in four panels, but in this case only for a total range of a factor of 10 in $n_H$ in units of cm$^{-3}$, from 0.25 to $2.5 \times 10^{11}$, with iso-temperature surfaces from 7500 to 12,500 K.  This limited density range covers all of the observed ratios for the cases with $T \ge 8750$ K.  In this temperature range the model curves are very similar, in that at a fixed temperature the Pa$n$/{\pbeta} value is small for each $n$ when the density is low, and the whole decrement curve rises as density increases, with a higher rise at higher $n$. The rise of the decrement to a flatter shape with increasing density happens more slowly when $T \le 7500$ K, so that for $T$ between 5000 and 7500~K (not shown in the figure) the required range of densities to cover all the observations is somewhat higher, from $10^{11}$ to $10^{12}$ cm$^{-3}$.

The same range of densities is shown in the ratio-ratio plots, which extend over a wider range of temperature from 5000 K to 20,000 K. The behavior of each ratio-ratio relation is similar, with line ratios increasing with increasing density and higher temperatures corresponding to a higher value of  {\pgamma}$/${\pbeta} or {\bten}$/${\bgamma} for a given value of {\bgamma}$/${\pbeta}. In this figure the iso-density contours bend back to somewhat lower ratios for $T \le 7500$ K, again reflecting the need for somewhat higher densities at lower temperatures to reproduce a given ratio.

Several things are apparent when comparing the KF relations to the observed values. In the series decrement plots in Figure~\ref{f.KF_NH}, there is a well defined trend such that as  {\pgamma}$/${\pbeta} increases, so does Pa$n$/{\pbeta}, in agreement with the observations. This behavior is not seen in the Case B decrements in Figure~\ref{f.B_NH}. In the ratio-ratio plots in Figure~\ref{f.KF_ratios}, the upward slope toward higher ratios in both pairs of lines is aligned with the predicted behavior for line ratios increasing with increasing density, covering a span from $\sim$ $2 \times 10^{10}$ to  $2 \times 10^{11}$ cm$^{-3}$ for $T\ge$ 8750~K, where stars with higher accretion rates have densities at the higher end of this range. Moreover, in contrast to Case B, here the same conclusions are reached from all three diagnostic relations.  All indicate the density in the hydrogen line formation region spans an order of magnitude and increases with increasing mass accretion rate.  Temperatures are not well constrained, since, as seen in the ratio-ratio plots, reddening vectors cut the closely spaced iso-temperature lines almost orthogonally and errors in $A_V$ of $1-2$ magnitudes span the full range of possible temperatures. However the densities are robustly determined and are not sensitive to extinction uncertainties.

Thus, in contrast to Case B, the KF local line excitation calculations give consistent results across all diagnostics for the density, and suggest densities in the hydrogen line formation region are similar among the CTTS in our sample, on the order of $\log n_H = 11$. There is also a suggestion that higher veiling stars on average have densities higher than low veiling stars by about a factor of $5-10$.  The temperature is not as readily identified, since uncertainties in extinction result in ratios that can intersect any of the temperature contours between 5000 and 20,000 K. In contrast to the much lower temperatures inferred from comparisons of line ratios with Case B \citep{bary08}, this range corresponds to temperatures where collisional excitation as a means of photon production is important. 

\subsection{Role of Extinction}

\begin{figure}
\includegraphics[width=\hsize] {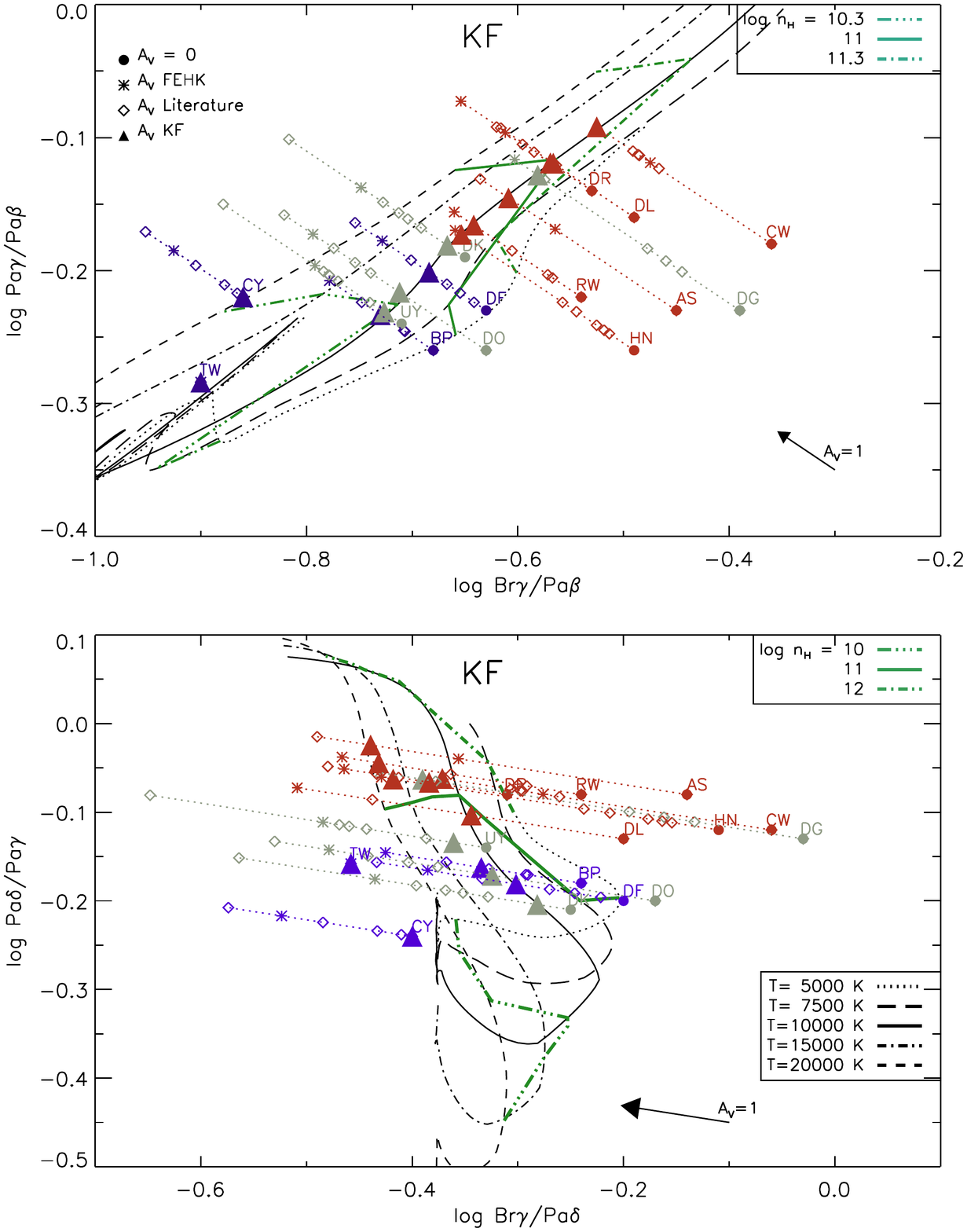}
\caption{Ratio-ratio plots of KF relations plus data, illustrating the range of published extinctions. Upper is {\pgamma}$/${\pbeta} versus {\bgamma}$/${\pbeta} and lower is {\pgamma}$/${\pdelta} versus {\bgamma}$/${\pdelta}. In both panels dotted lines connect ratios from the minimum and maximum reported $A_V$ for each star, where the minimum ($\bullet$) is the ratio uncorrected for extinction. Other points correspond to:\ $A_V$ from \citet{fis11}  ($*$),  $A_V$'s from the literature ($\diamond$), and the KF $A_V$ ($\Delta$).  In both cases, extrema of the literature values are outside the range of the model. Both use the same temperature legend, but a wider range of densities is shown in the lower panel.
\label{f.ext}}
\end{figure}

The previous figures applied the $A_V$ from FEHK to convert equivalent width ratios to intensity ratios. However, the range in $A_V$ reported in the literature is considerable so in the upper panel of Figure~\ref{f.ext} we again show the {\pgamma}$/${\pbeta} versus {\bgamma}$/${\pbeta}  relation comparing observations to the KF predictions, but this time show the `observed'  intensity ratio implied for each assessment of $A_V$ from the literature cited in Table~\ref{t.Av}, with from 2 to 7 $A_V$ per star. For comparison, we also include the directly observed ratio with no correction for extinction. In contrast to the minimal impact of the $A_V$ spread (up to 2.5 magnitudes for individual stars) on the relation between the accretion luminosity and the veiling shown in Figure~\ref{f.acc},  in a ratio-ratio plot the $A_V$ spread moves the `observed' ratios through a significant domain of model predictions in the direction of a reddening vector. Most apparent from the figure is that the extrema of the reported $A_V$ for each star often lie outside the range expected from either the KF or Case B relations for the range of densities and temperatures explored here. The FEHK values used here are also higher than the average value for most stars, however the largest values typically come from the study of \citet{fur11}, while the smallest values are from \citet{gul98, gul00} and \citet{ken95}. 

Even with the uncertainty in $A_V$ some conclusions can be drawn from comparing observed ratios to the KF calculations. First, the uncertainty in extinction does not alter the empirical conclusion that stars with higher veiling tend to have higher line ratios, since this trend is not affected by reddening vectors. In the KF models, the reddening vectors cross the iso-temperature lines orthogonally so that $n_H$ but not temperature is well constrained, with an implied range in density from $2 \times 10^{10}$ to $2 \times 10^{11}$ cm$^{-3}$, as noted in the last section.  The temperature is impossible to disentangle from the uncertainty in $A_V$; however, if we make the assumption that CTTS hydrogen lines form in a narrow temperature range then the spread in observed ratios across iso-temperature lines in the ratio-ratio plots could be due wholly to uncertainty in the adopted extinction for each star. 

Adopting this assumption with the KF predictions we can recover an $A_V$ that would place each star along a single temperature locus. While no such temperature assessment exists from first principles, $T \sim$ 10,000 K is a reasonable choice for both the magnetospheric accretion models of \citet{muz98a} and \citet{kur11} and the wind models of \citet{kwan11}.  We thus calculate a new $A_V$ for each star, indicated by filled triangles in Figure~\ref{f.ext} and listed in Column 6 of Table~\ref{t.Av}, corresponding to the extinction correction required to bring each star to the $T=$ 10,000~K iso-temperature line from the KF models.  This `KF $A_V$' is in the mid-range of previously determined values for most stars, and is on average 0.7 magnitudes smaller than the $A_V$ adopted throughout this paper from the FEHK study. For TW Hya the `KF $A_V$' is zero since it already lies on the $T=$ 10,000~K iso-temperature line. This is also the extinction that is found in all studies of this nearby star. This technique cannot be applied to CY Tau, since the uncorrected ratios lie on the $T=$ 20,000~K iso-temperature line. Thus although we identify `KF $A_V$'  = 0 for this star, this is not in line with previous estimates and likely means the uncorrected ratios are in error due to the large uncertainty in the definition of the {\bgamma} profile.  

In contrast if the same spread of implied ratios were plotted on the Case B ratio-ratio plot of Figure 8 the result would be ambiguous in inferring both density and temperature. The spread of up to 2.5 magnitudes in $A_V$ for each star means that reddening vectors would cross multiple iso-density lines orthogonally, and each iso-temperature surface would also be intersected twice, corresponding to a different implied density in each case. Under this scenario, even if one were to adjust the $A_V$ to locate all the stars along the same iso-density line, temperatures from 1000 to 20,000 K would be required to match the observations, and TW Hya would be several orders of magnitude higher density than the other stars. It is difficult to imagine that there is such a variety of physical conditions in the region where hydrogen lines form in CTTS, making the KF assumptions seem far more plausible than any Case B scenario.   

\begin{figure}
\includegraphics[angle=90,width=\hsize] {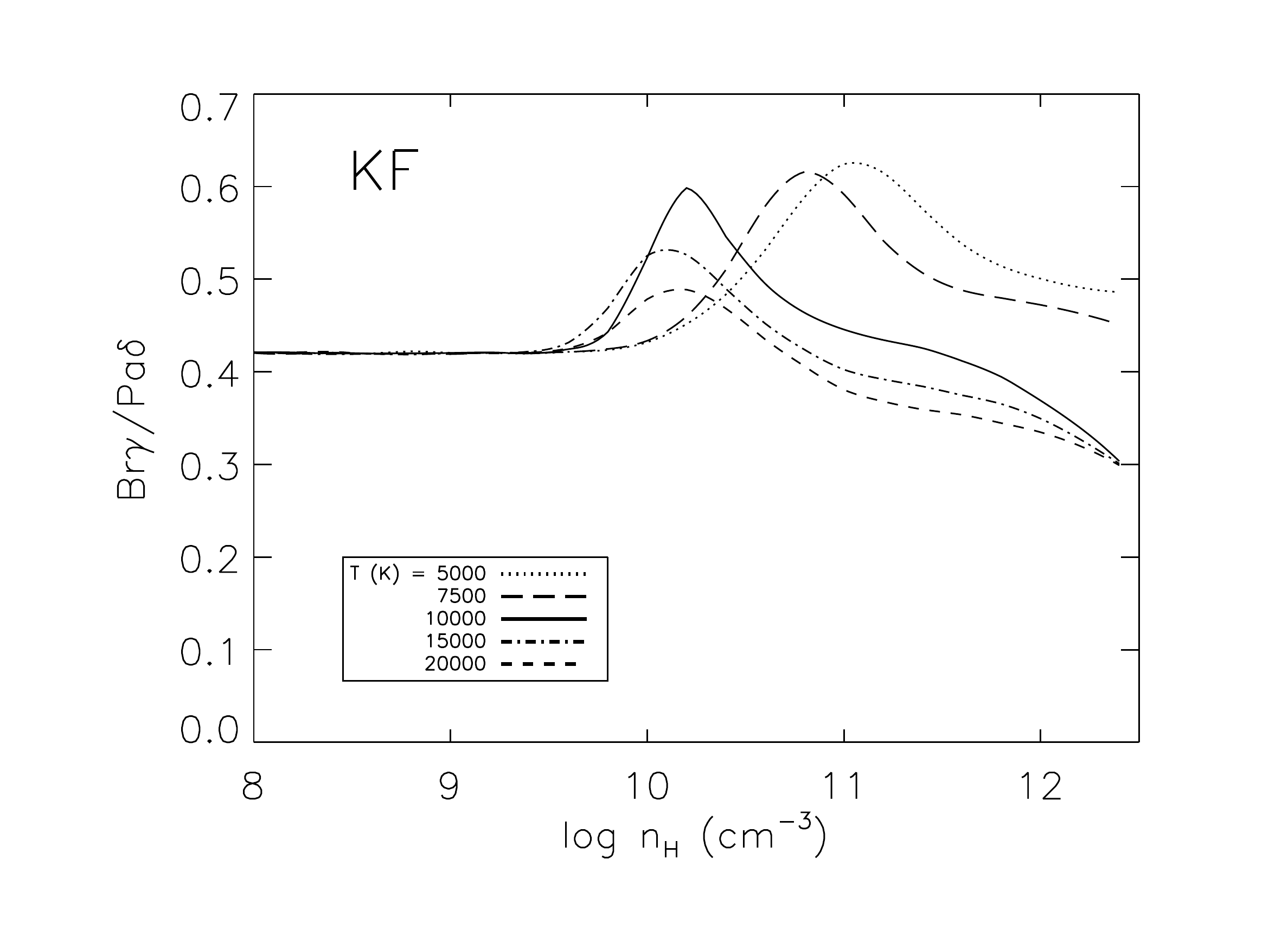}
\caption{The KF ratio of {\bgamma}$/${\pdelta} as a function of log $n_H$. Since  both transitions arise from the same $n=7$ upper level, even with a wide range of optical depths, this ratio is much less sensitive to density and has a smaller dynamic range than other ratios shown in Fig.\ 6.
\label{f.pdel}}
\end{figure}

Additional insight on extinction can be gleaned by turning to line ratios that arise from the same upper level, so the effects of density and temperature on the ratio are much reduced. To this end, we show in the lower panel of Figure~\ref{f.ext} a ratio-ratio plot of {\pdelta}$/${\pgamma} versus {\bgamma}$/${\pdelta}. The behavior of the KF calculations for this ratio-ratio relationship is different from those shown previously because both {\bgamma} and {\pdelta} arise from the same upper level ($n=7$). This ratio is shown as a function of $n_H$ and temperature in the KF calculations in Figure~\ref{f.pdel}, analogous to the ratios shown in Figure~\ref{f.theory}. However, this ratio behaves quite differently from the other line ratios studied here,  which rise continuously as the optical depth increases until reaching a plateau value. Instead, {\bgamma}$/${\pdelta} departs from the optically thin ratio around $n_H$ = $10^{10}$ cm$^{-3}$, first rising as {\pdelta} becomes optically thick, and then falling as the optical depths of both lines continue to increase. When included in a ratio-ratio plot, such as {\pdelta}$/${\pgamma} versus {\bgamma}$/${\pdelta}, the former ratio traces density while the latter is sensitive mostly to extinction, with a dynamic range about half that of the other line ratios over the same span of density and temperature. 

Returning the to lower panel of Figure~\ref{f.ext}, we have included the corresponding set of `observed' ratios for the same literature values of  $A_V$ shown in the upper panel. Again, the largest values of literature $A_V$ are outside the range of the models and the high veiling stars require an $A_V$ in excess of unity. For two thirds of the stars the ratios resulting from the `KF $A_V$', set by the $T=$ 10,000~K iso-temperature locus in the ratio-ratio plot of {\pgamma}$/${\pbeta} versus {\bgamma}$/${\pbeta} in Figure~\ref{f.ext}, fall on the same locus here suggesting it may offer a reasonable estimate for the extinction in many cases. 

Clearly there is a need for a more definitive assessment of extinction in T Tauri stars but that is not the purpose of this work. We do however conclude that the highest values of $A_V$ reported in the literature seem to be out of bounds for many stars. Similarly some of the low extinctions in the literature are incompatible with either the Case B or KF model for the five high mass accretion rate stars with strong emission excess at all wavelengths noted by FEHK (CW Tau, DG Tau, AS 353A, HN Tau and RW Aur). For these stars, A$_V$ in excess of 1 magnitude is the minimum required to move the line ratios into the realm of the models.  While we do not consider our `KF $A_V$' values to be definitive, since the assumption that the temperature in the hydrogen line formation region is identical in all CTTS is  an oversimplification, it will be of interest to compare them to new assessments of $A_V$ that will be possible with the new generation of high resolution and broad wavelength coverage spectrographs.  

\subsection{Extinction to Embedded Sources}

Correcting for extinction is notoriously difficult for embedded Class I protostars, when the $H$ and $K$ bands may be the shortest accessible wavelengths and scattered light from envelopes as well as emission from accretion related phenomena are present. Often a variety of techniques to recover $A_V$ are employed, and the result for a single star can differ by many magnitudes   \citep{bec07,con10,davis11,cog12}.  The near infrared ratio-ratio relations investigated here offer an alternate way of addressing this problem. While the locus of theoretical predictions in the ratio-ratio planes are similar in both Case B and the KF calculations, we have shown in previous sections that those from KF provide a more likely scenario for the observed line ratios and we thus focus this discussion on the KF predictions. If an observed pair of hydrogen line ratios, such as Pa$\gamma$/Pa$\beta$ versus Br$\gamma$/Pa$\beta$, or for more embedded objects Br10/Br$\gamma$ versus Br$\gamma$/Pa$\beta$, are tracked back along a reddening vector until they intersect the KF model relations, an $A_V$ can be recovered if a temperature is adopted. However, since the range of iso-temperature loci from 5000 K to 20,000 K span only two magnitudes along a reddening vector, any adopted temperature in that range will give $A_V$ with a fractional uncertainty that is modest for deeply embedded sources.

\begin{figure}
\includegraphics[width=\hsize] {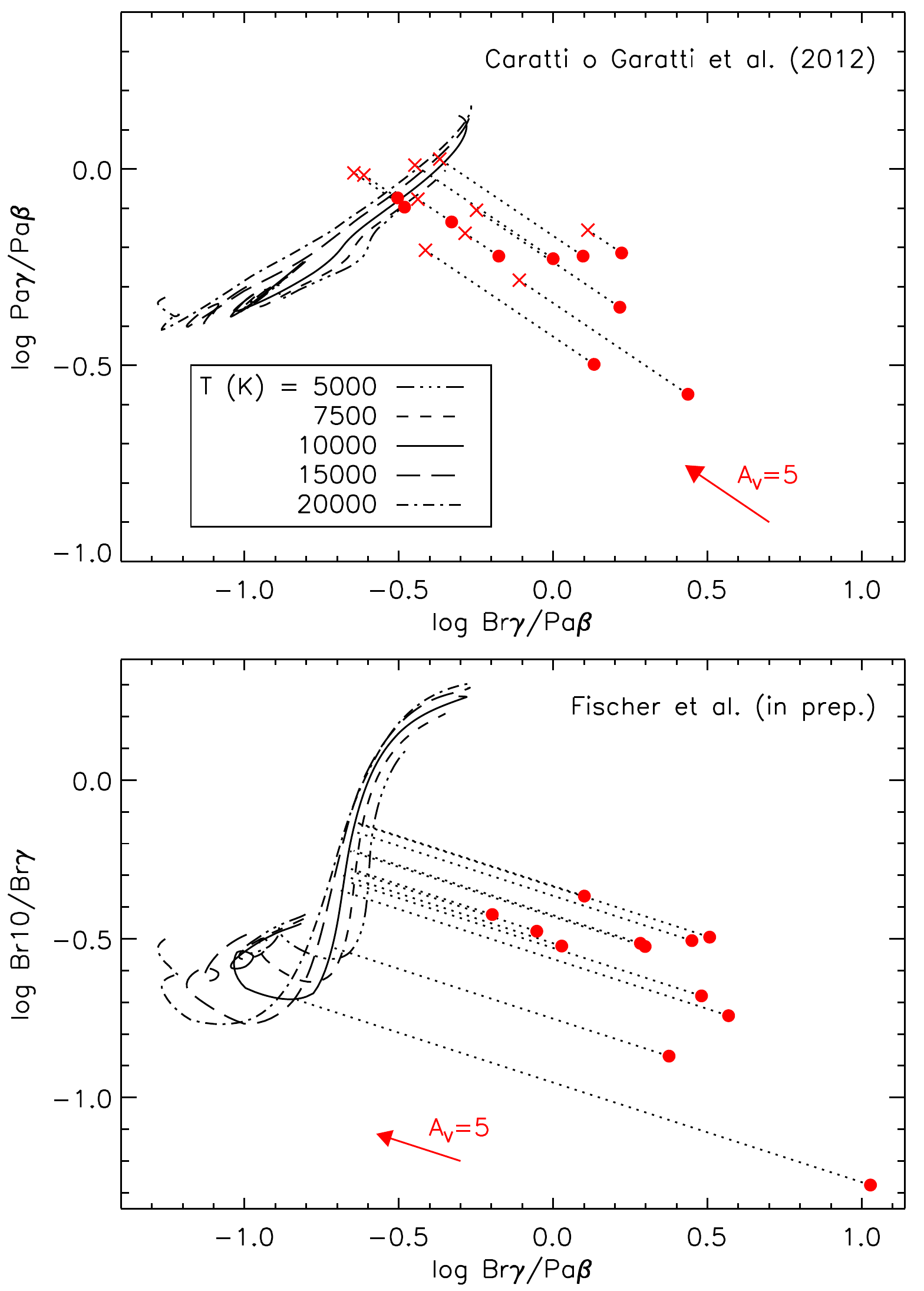}
\caption{Upper:\ Ratio-ratio plot of {\pgamma}$/${\pbeta} versus {\bgamma}$/${\pbeta} for 10 sources in L 1641 \citep{cog12}, plus the KF relations. Both directly observed ($\bullet$) and extinction corrected ($\times$) values are shown, using extinctions adopted by the authors. Lower:\ Ratio-ratio plot of {Br10}$/${\bgamma} versus {\bgamma}$/${\pbeta} for 12 deeply embedded Class I sources in Orion and the KF relations. Reddening vectors extend from the directly observed values to the KF $T=$ 10,000 K locus. In both panels, the density increases along the model tracks from lower left to upper right, as in Figure 10, except over a wider range from $n_H= 10^{8}$ to $2.5 \times 10^{12}$ cm$^{-3}$. \label{f.ext2}}
 \end{figure}

As an illustration, in the top panel of Figure~\ref{f.ext2} we show the ratio-ratio relation of {\pgamma}$/${\pbeta} versus {\bgamma}$/${\pbeta} for the KF calculations along with observations for the ten objects in L1641 for which intensities for all three lines were reported in a recent study by \citet{cog12}. The study focussed on a mix of Class I and II sources observed with NTT SOFI. We show the observed ratios for these ten objects, both uncorrected for extinction, and corrected with the extinction adopted by the authors, assessed from the mean of up to six different $A_V$ estimates after rejecting outlying values. If the reddening vectors are extended back from the uncorrected ratios to the $T=$ 10,000 K line in the KF relations, they would extend well beyond the adopted ratios in half of the objects, suggesting that the adopted $A_V$, which range from 2 to 10, may be significantly underestimated in some sources. 

As a second illustration, we show in the lower panel of Figure~\ref{f.ext2} the ratio-ratio relation of {\bten}$/${\bgamma} versus {\bgamma}$/${\pbeta} for the KF calculations along with observations of 12 deeply embedded protostars in Orion observed  with SpeX as part of the HOPS open-time key program of the Herschel Space Observatory \citep{fis13,man13,stu13}. The uncorrected line ratios for these protostars, a subset of Class I sources with $K<12.5$~mag identified from 2MASS and Spitzer colors between 1.2 and 24 {\micron} by \citet{meg12}, show that the extinctions are considerably higher than the objects from Caratti o Garrati. (There are no objects in common in the two samples.) Using the technique of  extending reddening vectors from the uncorrected ratios to the $T=$ 10,000 K line in the KF relations, the `KF $A_V$' fall between 10 and 34 magnitudes. The SEDs for these objects are being analyzed by the HOPS team, who determine extinctions by comparing the 1 to 870 {\micron} spectra and photometry of each object to a grid of radiative transfer models first described by \citet{ali10} and generated with the code of \citet{whi03}, which yields the extinction from the intervening interstellar dust, the circumstellar envelope, and the accretion disk. As will be shown in a forthcoming paper (Fischer et al., in prep.), the extinction derived in these two approaches is comparable, suggesting both may be a viable means of determining extinction to embedded sources, in contrast to commonly used approaches for Class I extinctions.

In these illustrations not only can an $A_V$ be recovered from the intersection of the reddening vectors with the loci of the KF iso-temperature calculations, but this intersection also implies a unique value of the hydrogen number density, $n_H$. For the 10 sources in L1641 the implied densities fall in the range $10^{11} -10^{12}$ cm$^{-3}$, somewhat higher than for the CTTS studied here. For the 12 sources from across the Orion complex, the implied densities span a wider range, from $10^{10} -10^{12}$ cm$^{-3}$. While further investigation of the validity of the KF models applied to accreting systems is required, we consider this a promising approach to further our understanding of Class I sources as well as CTTS.

\section{Discussion}

Paschen and Brackett lines are some of the strongest lines in near infrared spectra of accreting young stars. They are superior to the Balmer series for line ratio diagnostics since they are less subject to uncertainties in extinction, and to the opacity effects that riddle the Balmer series with blueshifted and redshifted absorption. We have focussed here on velocity resolved Paschen and Brackett lines in a small sample of classical T Tauri stars, primarily in Tau-Aur, in order to look for variations in the hydrogen line ratios as a function of mass accretion rate and to compare them to theoretical models for hydrogen line formation. The primary strength of this work  is to show the limitations of using Case B predictions to interpret CTTS hydrogen line ratios and to demonstrate the potential of using the \citet{kwan11} local line excitation predictions to infer physical conditions and extinction. 

We experimented with three methods for determining hydrogen line intensity ratios from emission equivalent widths. For the majority of our sample, those with strong emission and no redshifted absorption, we found almost identical ratios among the three methods. However, when the emission equivalent width is weak and/or if redshifted absorption is present, the line ratios corrected for these effects can be altered from directly observed values. For example in three stars the directly observed ratio of {\pgamma}$/${\pbeta} exceeded unity but dropped to $\sim0.8$  after correcting the directly measured line profile for the  underlying veiled photospheric absorption and taking ratios only in regions of the line profile free of redshifted absorption.  

While the ratio of {\pgamma}$/${\pbeta} can exceed unity in the KF models at densities exceeding $n_H$ = 10$^{12}$ cm$^{-3}$,  the ratio of {\bgamma}$/${\pbeta} is always below unity. Thus it is surprising that  while half the stars in a low resolution study of Chameleon I and II were found by \citet{ant} to have {\bgamma}$/${\pbeta} ratios similar to objects in Tau-Aur (0.1 to 0.4),  the remainder had larger ratios, with {\bgamma}$/${\pbeta} between 0.5 and 2. Such large ratios are inconsistent with any Case B scenario and outside of the range of KF calculations as well.  As noted by the authors, these ratios could be explained by invoking line formation in optically thick (i.e., blackbody) LTE emission with $T < 5000$ K.  However, the {\bgamma} intensity could appear to exceed {\pbeta} at low resolution if the lines are subject to redshifted absorption, which would reduce emission from {\pbeta} relative to {\bgamma}, and might go unnoticed in unresolved lines.  High resolution spectra of the Chamaeleon objects are required to see whether this group of stars has near infrared line profiles that differ from all other CTTS observed to date, which are generally similar in their kinematic properties and thus suggestive of formation under similar conditions \citep{muz98b,folha,edw06}.

After correcting the intensity ratios for the effects mentioned above, we found a loose relation between near infrared hydrogen line ratios and $Y$-band veiling, and thus the implied mass accretion rate. In the KF predictions, which we favor over Case B, this would correspond to somewhat higher densities in the hydrogen line formation region in stars with higher accretion rates. In a previous study we found that {\pbeta} line profiles from NIRSPEC at $R= $ 25,000 showed kinematic behavior that correlated with the $Y$-band veiling, where stars with higher veiling had broader line profiles \citep{edw06}. Taken together, these results suggest stars with higher disk accretion rates have both higher densities and higher velocities in the line formation region. These high veiling stars also show extended blue wings with velocities in their hydrogen profiles in excess of what can be produced in magnetospheric infall, so winds along with funnel flows are likely implicated. A larger sample is required to see whether this connection between veiling and line ratios is real or is simply due to our small sample size, and whether there is a connection to the profile morphology.

Both Case B and KF local line excitations predict a locus of line ratios that overlap with observations of CTTS; however, the behavior of the line ratio diagnostics leads us to favor the KF predictions over those of Case B. The inconsistencies in interpreting line ratios among different Case B diagnostics likely arises because the condition that the neutral hydrogen column density, $n_{HI} \delta l$, is sufficiently small that radiative de-excitation occurs more rapidly than collisional excitation from $n=2$, required to keep the Paschen and Brackett lines optically thin, is violated. Additional evidence for the latter is the fact that 24\% of CTTS show redshifted absorption in {\pgamma} when examined at high spectral resolution \citep{edw06}, which requires line opacities well in excess of unity. In contrast the KF predictions cover a large range of possible line opacities, self consistently taking into account the local density, temperature, and ionization rate. To enable others to compare their data with these calculations, we have assembled a set of hydrogen line ratios from  the KF local line excitation models and make them publicly available on a web server for other researchers acquiring near infrared spectra of CTTS\footnote{Currently at http://earth.ast.smith.edu/sedwards/KFweb/, files of the model output will be available in a revised format at the journal website by the time of publication.}. The site includes both the ratios used here, plus others based on H$\alpha$, Br$\alpha$ (4.05 $\mu$m), H7--6 (12.4 $\mu$m) and H9--7 (11.3 $\mu$m) that may be useful for existing or planned observations. 

Under the KF assumptions, the implied density in the hydrogen line formation region is within a factor of a few times $10^{11}$ cm$^{-3}$ for the 16 CTTS in our sample, with the higher accretion rate stars at the higher end of the range. Here we compare these densities to those expected from hydrogen lines that are formed in accretion funnels, and find they are somewhat lower than expected based on the models of \citet{muz98a}, also adopted by \citet{kur11}. In these models  Balmer, Paschen and Brackett emission arises over the full length of the accretion columns and the mass accretion rate sets the density in the accretion columns.  For example, in Figure 2 of \citet{muz98a}, for a fiducial case of an aligned, symmetric dipole flow with a maximum temperature of 8000 K and ${\dot M}_{\rm acc}$ = $10^{-7}$ {\msunyr},  the density increases from $10^{12}$ to $10^{13}$ cm$^{-3}$ along the region of the accretion column where infall velocities exceed 100~{\kms} and much of the line emission arises. This ${\dot M}_{\rm acc}$ is in line with about half the stars in our sample, based on {\pbeta} line luminosities and shown in Figure 1, yet the  hydrogen number densities we infer for these stars are at least an order of magnitude lower. For the remainder of our sample, mass accretion rates are between ${\dot M}_{\rm acc}= 10^{-9}$ and $10^{-8}$ {\msunyr}, which would have correspondingly lower densities in the funnel flow model. Thus although the general behavior expected in the funnel flow models, that higher accretion rate objects would have higher densities in the accretion columns, is born out in the observations, there are discrepancies.  Whether these inconsistencies are serious challenges to the common assumption that hydrogen lines are formed primarily in funnel flows or not is difficult to say at this stage of the investigation.

Another approach is to note that in magnetospheric accretion models the density in the immediate pre-shock gas is about an order of magnitude higher than that over most of the flow due to the channeling effect of the `funnel' (e.g., figures in \citealt{muz98a} and \citealt{kur11}). Thus if we amplify the number density we infer in the hydrogen line formation region, $\sim 10^{11}$ cm$^{-3}$,  by an order of magnitude to $\sim 10^{12}$ cm$^{-3}$ to reflect the corresponding immediate pre-shock densities, we can compare this to pre-shock densities based on modeling either optical/UV continuum emission excess or line ratios in the X-ray domain. For example, \citet{cal98}  define an energy flux $F$ carried into the accretion shock by the funnel flow, where $F = 0.5\rho v_s^3$. In their models the free fall velocity $v_s$ is kept constant at around 300 {\kms} and the density $\rho $ in the immediate pre-shock gas sets the accretion shock energy.  The correspondence between  ${\dot M}_{\rm acc}$ and $\log F$ is given by their Equation 11. Applying this with the same ${\dot M}_{\rm acc}$ as in the fiducial case cited above, ${\dot M}_{\rm acc} = 10^{-7}$ {\msunyr}, the corresponding $F=10^{11}$ erg cm$^{-2}$ s$^{-1}$ implies a pre-shock number density  $\sim 10^{13}$ cm$^{-3}$, again about an order of magnitude higher than inferred from the near infrared line ratios. For stars at the low end of the mass accretion rates thought to apply to our sample, ${\dot M}_{\rm acc} = 10^{-9}$ {\msunyr}, the pre-shock number density would be two orders of magnitude smaller at $10^{11}$ cm$^{-3}$, in this case larger than we would infer for the pre-shock density of these stars.

We can also look at the pre-shock densities derived from modeling the He-like triplets of \ion{Ne}{9} and \ion{O}{7} in X-ray spectra. These line ratios in CTTS cannot be explained by coronal emission and are attributed to formation in the accretion shock \citep{gudel09}. A study of these line ratios in several low accretion rate CTTS indicate pre-shock densities $\sim 10^{13}$ cm$^{-3}$ for two stars in our current study, BP Tau and TW Hya \citep{gun11}, again an order of magnitude higher than we would infer.  Moreover, although these pre-shock densities are in line with those from magnetospheric accretion models, the inferred mass accretion rates based on x-ray line ratios are $10^{-9}$ {\msunyr} for BP Tau and $10^{-11}$ {\msunyr} for TW Hya \citep{gun11}, an order of magnitude or more lower than would be inferred from Figure 1 based on {\pbeta} luminosities or from the accretion shock models of \citet{cal98} based on optical/UV continuum excess.  The tendency for mass accretion rates inferred from X-ray line ratios to be consistently lower than those based on accretion shock models of optical/UV emission excess is recognized by \citet{gun11}, who suggest that this might arise from inhomogeneous spots, partial absorption in buried shocks, or the presence of accretion streams that impact at velocities considerably below free-fall speeds (see also \citealt{ingleby13}). 

In this paper we have applied the KF calculations only to hydrogen line ratios. The KF local line excitation calculations also include \ion{He}{1}, \ion{Ca}{2}, \ion{O}{1}, and \ion{Na}{1} so that when ratios of other species are included, we will be able not only to test the conclusions regarding density, but also set constraints on the temperature as well. Although the accretion funnel paradigm as a source of the near infrared hydrogen emission lines has withstood more than a decade of scrutiny, there are growing indications that a deeper investigation is warranted. In an era where it will now be possible to simultaneously match line luminosities, line ratios, and high resolution line profiles from multiple lines over a wide spectral range and over several stellar rotation periods to those predicted from modern magnetospheric accretion models \citep{long11, kur13}, with misaligned fields with multipole components, coupled with rigorous diagnostics of the physical conditions in the line formation region, we may be able to decipher what the relative contributions of funnel flows, accretion shocks, winds and the inner disk might be to the rich emission line spectra of CTTS.

\section{Conclusions}

The key conclusions from this work are
\begin{itemize}
\item{ In our limited sample, we see a tendency for the hydrogen line ratios in CTTS to have different behavior in stars of high and low veiling. In the higher veiling stars the Paschen decrement is shallower and line ratios tend to be higher in ratio-ratio relations such as {\pgamma}$/${\pbeta} versus {\bgamma}$/${\pbeta}  and {\bten}$/${\bgamma} versus {\bgamma}$/${\pbeta}. This conclusion is independent of extinction uncertainties. However it is based on a relatively small sample of stars and needs to be examined in a larger context.} 

\item{ The \citet{kwan11} local line excitation calculations offer a more consistent interpretation of physical conditions in the hydrogen line formation region of T Tauri stars than Case B. Under KF assumptions the density in the hydrogen line formation region lies within the range $n_H = 2 \times 10^{10} - 2 \times 10^{11}$ cm$^{-3}$, with densities higher on average in stars with higher accretion rates. Extinction uncertainties preclude a reliable temperature determination. Under Case B assumptions not only do different diagnostics yield different implied values of temperature and electron density in CTTS, the range of implied values exceeds four orders of magnitude in electron density and a factor of ten in temperature. } 

\item{ The largest source of uncertainty in determining line ratios in T Tauri stars is in correcting for extinction. The range of $A_V$ values in the literature can be 2.5 magnitudes for a given star. This makes comparisons with line excitation models problematic for typical T Tauri stars, but for deeply embedded sources, where an error in $A_V$ of two magnitudes is less consequential,  the KF models can be used to evaluate both the extinction and the density in the line formation region when near infrared hydrogen line ratios can be determined. }
\end{itemize}

\acknowledgments Thanks to Greg Herczeg, Mike Peterson, and Hans Moritz G{\"u}nther for useful discussions. Also, thanks to our referee who made some very helpful comments. The authors reverently acknowledge the cultural significance of the Mauna Kea summit to the indigenous Hawaiian community.  We are most fortunate to have had the opportunity to conduct observations with IRTF from this mountain.

{\it Facilities:} \facility{IRTF(SpeX)}

\end{document}